\documentclass[longauth]{aa} 

\usepackage{graphicx}
\usepackage{txfonts}
\usepackage{tabularx}
\usepackage{multirow}
\usepackage{booktabs}
\usepackage[]{hyperref}
\hypersetup{
    colorlinks = true,
    linkcolor = {blue}, 
    citecolor = {blue},
    urlcolor = {blue}
}
\begin{document}

   \title{CHEOPS and TESS view of the ultra-short period super-Earth TOI-561\,b\thanks{Based in part on Guaranteed Time Observations on the European Space Agency telescope CHEOPS under programme 10.0008 by the CHEOPS consortium. The raw and detrended photometric time series data are available in electronic form at the CDS via anonymous ftp to cdsarc.cds.unistra.fr (130.79.128.5) or via \url{https://cdsarc.cds.unistra.fr/cgi-bin/qcat?J/A+A/}}}


\author{
J. A. Patel\inst{1}\thanks{\email{jayshil.patel@astro.su.se}} $^{\href{https://orcid.org/0000-0001-5644-6624}{\includegraphics[scale=0.5]{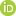}}}$, 
J. A. Egger\inst{2} $^{\href{https://orcid.org/0000-0003-1628-4231}{\includegraphics[scale=0.5]{figures/orcid.jpg}}}$, 
T. G. Wilson\inst{3,4} $^{\href{https://orcid.org/0000-0001-8749-1962}{\includegraphics[scale=0.5]{figures/orcid.jpg}}}$,
V. Bourrier\inst{5} $^{\href{https://orcid.org/0000-0002-9148-034X}{\includegraphics[scale=0.5]{figures/orcid.jpg}}}$,
L. Carone\inst{6}, 
M. Beck\inst{5} $^{\href{https://orcid.org/0000-0003-3926-0275}{\includegraphics[scale=0.5]{figures/orcid.jpg}}}$,
D. Ehrenreich\inst{5,7} $^{\href{https://orcid.org/https://www.notion.so/D-Ehrenreich-7e5007d8e19f46368c036b8dfbbd5a59}{\includegraphics[scale=0.5]{figures/orcid.jpg}}}$,
S. G. Sousa\inst{8} $^{\href{https://orcid.org/0000-0001-9047-2965}{\includegraphics[scale=0.5]{figures/orcid.jpg}}}$,
W. Benz\inst{2,9} $^{\href{https://orcid.org/0000-0001-7896-6479}{\includegraphics[scale=0.5]{figures/orcid.jpg}}}$,
A. Brandeker\inst{1} $^{\href{https://orcid.org/0000-0002-7201-7536}{\includegraphics[scale=0.5]{figures/orcid.jpg}}}$,
A. Deline\inst{5} $^{\href{https://orcid.org/https://www.notion.so/A-Deline-TS1-9b818a0c88e04c45910c1e49e0a4397d}{\includegraphics[scale=0.5]{figures/orcid.jpg}}}$,
Y. Alibert\inst{2} $^{\href{https://orcid.org/0000-0002-4644-8818}{\includegraphics[scale=0.5]{figures/orcid.jpg}}}$,
K. W. F. Lam\inst{10} $^{\href{https://orcid.org/0000-0002-9910-6088}{\includegraphics[scale=0.5]{figures/orcid.jpg}}}$,
M. Lendl\inst{5} $^{\href{https://orcid.org/0000-0001-9699-1459}{\includegraphics[scale=0.5]{figures/orcid.jpg}}}$,
R. Alonso\inst{11,12},
G. Anglada\inst{13,14} $^{\href{https://orcid.org/0000-0002-3645-5977}{\includegraphics[scale=0.5]{figures/orcid.jpg}}}$,
T. Bárczy\inst{15} $^{\href{https://orcid.org/0000-0002-7822-4413}{\includegraphics[scale=0.5]{figures/orcid.jpg}}}$,
D. Barrado\inst{16} $^{\href{https://orcid.org/0000-0002-5971-9242}{\includegraphics[scale=0.5]{figures/orcid.jpg}}}$,
S. C. C. Barros\inst{8,17} $^{\href{https://orcid.org/0000-0003-2434-3625}{\includegraphics[scale=0.5]{figures/orcid.jpg}}}$,
W. Baumjohann\inst{6} $^{\href{https://orcid.org/0000-0001-6271-0110}{\includegraphics[scale=0.5]{figures/orcid.jpg}}}$,
T. Beck\inst{2}, 
N. Billot\inst{5} $^{\href{https://orcid.org/0000-0003-3429-3836}{\includegraphics[scale=0.5]{figures/orcid.jpg}}}$,
X. Bonfils\inst{18} $^{\href{https://orcid.org/0000-0001-9003-8894}{\includegraphics[scale=0.5]{figures/orcid.jpg}}}$,
C. Broeg\inst{2,9} $^{\href{https://orcid.org/0000-0001-5132-2614}{\includegraphics[scale=0.5]{figures/orcid.jpg}}}$,
M.-D. Busch\inst{2}, 
J. Cabrera\inst{10}, 
S. Charnoz\inst{19} $^{\href{https://orcid.org/0000-0002-7442-491X}{\includegraphics[scale=0.5]{figures/orcid.jpg}}}$,
A. Collier Cameron\inst{3} $^{\href{https://orcid.org/0000-0002-8863-7828}{\includegraphics[scale=0.5]{figures/orcid.jpg}}}$,
Sz. Csizmadia\inst{10} $^{\href{https://orcid.org/0000-0001-6803-9698}{\includegraphics[scale=0.5]{figures/orcid.jpg}}}$,
M. B. Davies\inst{20} $^{\href{https://orcid.org/0000-0001-6080-1190}{\includegraphics[scale=0.5]{figures/orcid.jpg}}}$,
M. Deleuil\inst{21} $^{\href{https://orcid.org/0000-0001-6036-0225}{\includegraphics[scale=0.5]{figures/orcid.jpg}}}$,
L. Delrez\inst{22,23} $^{\href{https://orcid.org/0000-0001-6108-4808}{\includegraphics[scale=0.5]{figures/orcid.jpg}}}$,
O. D. S. Demangeon\inst{8,17} $^{\href{https://orcid.org/0000-0001-7918-0355}{\includegraphics[scale=0.5]{figures/orcid.jpg}}}$,
B.-O. Demory\inst{9} $^{\href{https://orcid.org/0000-0002-9355-5165}{\includegraphics[scale=0.5]{figures/orcid.jpg}}}$,
A. Erikson\inst{10}, 
A. Fortier\inst{2,9} $^{\href{https://orcid.org/0000-0001-8450-3374}{\includegraphics[scale=0.5]{figures/orcid.jpg}}}$,
L. Fossati\inst{6} $^{\href{https://orcid.org/0000-0003-4426-9530}{\includegraphics[scale=0.5]{figures/orcid.jpg}}}$,
M. Fridlund\inst{24,25} $^{\href{https://orcid.org/0000-0002-0855-8426}{\includegraphics[scale=0.5]{figures/orcid.jpg}}}$,
D. Gandolfi\inst{26} $^{\href{https://orcid.org/0000-0001-8627-9628}{\includegraphics[scale=0.5]{figures/orcid.jpg}}}$,
M. Gillon\inst{22} $^{\href{https://orcid.org/0000-0003-1462-7739}{\includegraphics[scale=0.5]{figures/orcid.jpg}}}$,
M. Güdel\inst{27}, 
K. Heng\inst{9,4} $^{\href{https://orcid.org/0000-0003-1907-5910}{\includegraphics[scale=0.5]{figures/orcid.jpg}}}$,
S. Hoyer\inst{21} $^{\href{https://orcid.org/0000-0003-3477-2466}{\includegraphics[scale=0.5]{figures/orcid.jpg}}}$,
K. G. Isaak\inst{28} $^{\href{https://orcid.org/0000-0001-8585-1717}{\includegraphics[scale=0.5]{figures/orcid.jpg}}}$,
L. L. Kiss\inst{29,30}, 
E. Kopp\inst{31}, 
J. Laskar\inst{32} $^{\href{https://orcid.org/0000-0003-2634-789X}{\includegraphics[scale=0.5]{figures/orcid.jpg}}}$,
A. Lecavelier des Etangs\inst{33} $^{\href{https://orcid.org/0000-0002-5637-5253}{\includegraphics[scale=0.5]{figures/orcid.jpg}}}$,
C. Lovis\inst{5} $^{\href{https://orcid.org/0000-0001-7120-5837}{\includegraphics[scale=0.5]{figures/orcid.jpg}}}$,
D. Magrin\inst{34} $^{\href{https://orcid.org/0000-0003-0312-313X}{\includegraphics[scale=0.5]{figures/orcid.jpg}}}$,
P. F. L. Maxted\inst{35} $^{\href{https://orcid.org/0000-0003-3794-1317}{\includegraphics[scale=0.5]{figures/orcid.jpg}}}$,
V. Nascimbeni\inst{34} $^{\href{https://orcid.org/0000-0001-9770-1214}{\includegraphics[scale=0.5]{figures/orcid.jpg}}}$,
G. Olofsson\inst{1} $^{\href{https://orcid.org/0000-0003-3747-7120}{\includegraphics[scale=0.5]{figures/orcid.jpg}}}$,
R. Ottensamer\inst{27}, 
I. Pagano\inst{36} $^{\href{https://orcid.org/0000-0001-9573-4928}{\includegraphics[scale=0.5]{figures/orcid.jpg}}}$,
E. Pallé\inst{11} $^{\href{https://orcid.org/0000-0003-0987-1593}{\includegraphics[scale=0.5]{figures/orcid.jpg}}}$,
G. Peter\inst{31} $^{\href{https://orcid.org/0000-0001-6101-2513}{\includegraphics[scale=0.5]{figures/orcid.jpg}}}$,
G. Piotto\inst{34,37} $^{\href{https://orcid.org/0000-0002-9937-6387}{\includegraphics[scale=0.5]{figures/orcid.jpg}}}$,
D. Pollacco\inst{4}, 
D. Queloz\inst{38,39} $^{\href{https://orcid.org/0000-0002-3012-0316}{\includegraphics[scale=0.5]{figures/orcid.jpg}}}$,
R. Ragazzoni\inst{34,37} $^{\href{https://orcid.org/0000-0002-7697-5555}{\includegraphics[scale=0.5]{figures/orcid.jpg}}}$,
N. Rando\inst{40}, 
F. Ratti\inst{40}, 
H. Rauer\inst{10,41,42} $^{\href{https://orcid.org/0000-0002-6510-1828}{\includegraphics[scale=0.5]{figures/orcid.jpg}}}$,
I. Ribas\inst{13,14} $^{\href{https://orcid.org/0000-0002-6689-0312}{\includegraphics[scale=0.5]{figures/orcid.jpg}}}$,
N. C. Santos\inst{8,17} $^{\href{https://orcid.org/0000-0003-4422-2919}{\includegraphics[scale=0.5]{figures/orcid.jpg}}}$,
G. Scandariato\inst{36} $^{\href{https://orcid.org/0000-0003-2029-0626}{\includegraphics[scale=0.5]{figures/orcid.jpg}}}$,
D. Ségransan\inst{5} $^{\href{https://orcid.org/0000-0003-2355-8034}{\includegraphics[scale=0.5]{figures/orcid.jpg}}}$,
A. E. Simon\inst{2} $^{\href{https://orcid.org/0000-0001-9773-2600}{\includegraphics[scale=0.5]{figures/orcid.jpg}}}$,
A. M. S. Smith\inst{10} $^{\href{https://orcid.org/0000-0002-2386-4341}{\includegraphics[scale=0.5]{figures/orcid.jpg}}}$,
M. Steller\inst{6} $^{\href{https://orcid.org/0000-0003-2459-6155}{\includegraphics[scale=0.5]{figures/orcid.jpg}}}$,
Gy. M. Szabó\inst{43,44}, 
N. Thomas\inst{2}, 
S. Udry\inst{5} $^{\href{https://orcid.org/0000-0001-7576-6236}{\includegraphics[scale=0.5]{figures/orcid.jpg}}}$,
B. Ulmer\inst{31}, 
V. Van Grootel\inst{23} $^{\href{https://orcid.org/0000-0003-2144-4316}{\includegraphics[scale=0.5]{figures/orcid.jpg}}}$,
V. Viotto\inst{34}, 
N. A. Walton\inst{45} $^{\href{https://orcid.org/0000-0003-3983-8778}{\includegraphics[scale=0.5]{figures/orcid.jpg}}}$}

\authorrunning{Patel et al.}


\institute{
\label{inst:1} Department of Astronomy, Stockholm University, AlbaNova University Center, 10691 Stockholm, Sweden \and
\label{inst:2} Physikalisches Institut, University of Bern, Gesellschaftsstrasse 6, 3012 Bern, Switzerland \and
\label{inst:3} Centre for Exoplanet Science, SUPA School of Physics and Astronomy, University of St Andrews, North Haugh, St Andrews KY16 9SS, UK \and
\label{inst:4} Department of Physics, University of Warwick, Gibbet Hill Road, Coventry, CV4 7AL, UK \and
\label{inst:5} Observatoire Astronomique de l'Universit\'e de Gen\`eve, Chemin Pegasi 51, CH-1290 Versoix, Switzerland \and
\label{inst:6} Space Research Institute, Austrian Academy of Sciences, Schmiedlstrasse 6, A-8042 Graz, Austria \and
\label{inst:7} Centre Vie dans l'Univers, Facult\'e des sciences, Universit\'e de Gen\`eve, Quai Ernest-Ansermet 30, CH-1211 Gen\`eve 4, Switzerland \and
\label{inst:8} Instituto de Astrofisica e Ciencias do Espaco, Universidade do Porto, CAUP, Rua das Estrelas, 4150-762 Porto, Portugal \and
\label{inst:9} Center for Space and Habitability, University of Bern, Gesellschaftsstrasse 6, 3012 Bern, Switzerland \and
\label{inst:10} Institute of Planetary Research, German Aerospace Center (DLR), Rutherfordstrasse 2, 12489 Berlin, Germany \and
\label{inst:11} Instituto de Astrofisica de Canarias, 38200 La Laguna, Tenerife, Spain \and
\label{inst:12} Departamento de Astrofisica, Universidad de La Laguna, 38206 La Laguna, Tenerife, Spain \and
\label{inst:13} Institut de Ciencies de l'Espai (ICE, CSIC), Campus UAB, Can Magrans s/n, 08193 Bellaterra, Spain \and
\label{inst:14} Institut d'Estudis Espacials de Catalunya (IEEC), 08034 Barcelona, Spain \and
\label{inst:15} Admatis, 5. Kandó Kálmán Street, 3534 Miskolc, Hungary \and
\label{inst:16} Depto. de Astrofisica, Centro de Astrobiologia (CSIC-INTA), ESAC campus, 28692 Villanueva de la Cañada (Madrid), Spain \and
\label{inst:17} Departamento de Fisica e Astronomia, Faculdade de Ciencias, Universidade do Porto, Rua do Campo Alegre, 4169-007 Porto, Portugal \and
\label{inst:18} Université Grenoble Alpes, CNRS, IPAG, 38000 Grenoble, France \and
\label{inst:19} Université de Paris, Institut de physique du globe de Paris, CNRS, F-75005 Paris, France \and
\label{inst:20} Centre for Mathematical Sciences, Lund University, Box 118, 221 00 Lund, Sweden \and
\label{inst:21} Aix Marseille Univ, CNRS, CNES, LAM, 38 rue Frédéric Joliot-Curie, 13388 Marseille, France \and
\label{inst:22} Astrobiology Research Unit, Université de Liège, Allée du 6 Août 19C, B-4000 Liège, Belgium \and
\label{inst:23} Space sciences, Technologies and Astrophysics Research (STAR) Institute, Université de Liège, Allée du 6 Août 19C, 4000 Liège, Belgium \and
\label{inst:24} Leiden Observatory, University of Leiden, PO Box 9513, 2300 RA Leiden, The Netherlands \and
\label{inst:25} Department of Space, Earth and Environment, Chalmers University of Technology, Onsala Space Observatory, 439 92 Onsala, Sweden \and
\label{inst:26} Dipartimento di Fisica, Universita degli Studi di Torino, via Pietro Giuria 1, I-10125, Torino, Italy \and
\label{inst:27} Department of Astrophysics, University of Vienna, Türkenschanzstrasse 17, 1180 Vienna, Austria \and
\label{inst:28} Science and Operations Department - Science Division (SCI-SC), Directorate of Science, European Space Agency (ESA), European Space Research and Technology Centre (ESTEC),
Keplerlaan 1, 2201-AZ Noordwijk, The Netherlands \and
\label{inst:29} Konkoly Observatory, Research Centre for Astronomy and Earth Sciences, 1121 Budapest, Konkoly Thege Miklós út 15-17, Hungary \and
\label{inst:30} ELTE E\"otv\"os Lor\'and University, Institute of Physics, P\'azm\'any P\'eter s\'et\'any 1/A, 1117 \and
\label{inst:31} Institute of Optical Sensor Systems, German Aerospace Center (DLR), Rutherfordstrasse 2, 12489 Berlin, Germany \and
\label{inst:32} IMCCE, UMR8028 CNRS, Observatoire de Paris, PSL Univ., Sorbonne Univ., 77 av. Denfert-Rochereau, 75014 Paris, France \and
\label{inst:33} Institut d'astrophysique de Paris, UMR7095 CNRS, Université Pierre \& Marie Curie, 98bis blvd. Arago, 75014 Paris, France \and
\label{inst:34} INAF, Osservatorio Astronomico di Padova, Vicolo dell'Osservatorio 5, 35122 Padova, Italy \and
\label{inst:35} Astrophysics Group, Keele University, Staffordshire, ST5 5BG, United Kingdom \and
\label{inst:36} INAF, Osservatorio Astrofisico di Catania, Via S. Sofia 78, 95123 Catania, Italy \and
\label{inst:37} Dipartimento di Fisica e Astronomia "Galileo Galilei", Universita degli Studi di Padova, Vicolo dell'Osservatorio 3, 35122 Padova, Italy \and
\label{inst:38} ETH Zurich, Department of Physics, Wolfgang-Pauli-Strasse 2, CH-8093 Zurich, Switzerland \and
\label{inst:39} Cavendish Laboratory, JJ Thomson Avenue, Cambridge CB3 0HE, UK \and
\label{inst:40} ESTEC, European Space Agency, 2201AZ, Noordwijk, NL \and
\label{inst:41} Zentrum für Astronomie und Astrophysik, Technische Universität Berlin, Hardenbergstr. 36, D-10623 Berlin, Germany \and
\label{inst:42} Institut für Geologische Wissenschaften, Freie Universität Berlin, 12249 Berlin, Germany \and
\label{inst:43} ELTE E\"otv\"os Lor\'and University, Gothard Astrophysical Observatory, 9700 Szombathely, Szent Imre h. u. 112, Hungary \and
\label{inst:44} MTA-ELTE Exoplanet Research Group, 9700 Szombathely, Szent Imre h. u. 112, Hungary \and
\label{inst:45} Institute of Astronomy, University of Cambridge, Madingley Road, Cambridge, CB3 0HA, United Kingdom}




   \date{Received XX; accepted XX}

 
  \abstract
   {Ultra-short period planets (USPs) are a unique class of super-Earths with an orbital period of less than a day and hence subject to intense radiation from their host star. These planets cannot retain a primordial H/He atmosphere, and most of them are indeed consistent with being bare rocky cores. A few USPs, however, show evidence for a heavyweight envelope, which could be a water layer resilient to evaporation, or a secondary metal-rich atmosphere sustained by outgassing of the molten, volcanic surface. Much thus remains to be learned about the nature and formation of USPs.} 
   {The prime goal of the present work is to refine the bulk planetary properties of the recently discovered TOI-561\,b, through the study of its transits and occultations. This is crucial to understand the internal structure of this USP and assess the presence of an atmosphere. }
   {We obtained ultra-precise transit photometry of TOI-561\,b with CHEOPS, and performed a joint analysis of this data along with three archival visits from CHEOPS and four TESS sectors.}
   {Our analysis of TOI-561\,b transit photometry put
   strong constraints on its properties. In particular, we restrict the uncertainties on the planetary radius at $\sim 2\%$ retrieving {\small $R_p = 1.42 \pm 0.02 \, R_\oplus$}. This result informs our internal structure modelling of the planet, which shows that the observations are consistent with negligible H/He atmosphere, however requiring other lighter materials, in addition to pure iron core and silicate mantle to explain the observed density. 
   We find that this can be explained by the inclusion of a water layer in our model. Additionally, we run a grid of forward models with a water-enriched atmosphere to explain the transit radius.
   We searched for variability in the measured $R_p/R_\star$ over time, which could trace changes in the structure of the planetary envelope. No temporal variations are however recovered within the present data precision.
   In addition to the transit event, we tentatively detect occultation signal in the TESS data with an eclipse depth {\small $L = 27.40 ^{+10.87} _{-11.35}$\,ppm}. We use models of outgassed atmospheres from the literature to explain this eclipse signal. We find that the thermal emission from the planet can mostly explain the observation. Based on this, we predict that near/mid-infrared observations with the James Webb Space Telescope should be able to detect silicate species in the atmosphere of the planet.
   This could also reveal important clues about the planetary interior and help disentangle planet formation and evolution models.}
   {}
   {}


   \maketitle
%

\section{Introduction}\label{sec:intro}
The advent of the new generation of exoplanet discovery missions, such as Kepler/K2 \citep{2010Sci...327..977B, 2014PASP..126..398H} and the Transiting Exoplanet Survey Satellite \citep[TESS;][]{2014SPIE.9143E..20R}, has led to the detection of the vast majority of known exoplanets. These planets and planetary systems show a large diversity in terms of composition and orbital architecture. In particular there exists one curious class of lower mass planets with radius smaller than $\sim 2 R_\oplus$ and that orbit their host stars within a day, and hence dubbed as ultra-short period planets (USPs) \citep[see,][ for a review]{2018NewAR..83...37W}. 
USPs receive extreme radiation from their host star, leading to the loss of any primary H/He atmosphere these planet may have formed with \citep{2013ApJ...776....2L, 2013ApJ...775..105O, 2016ApJ...817..107O}. These planets could, however, develop and sustain secondary atmospheres made up of heavier species. This may consist of a variety of species such as Na, K, SiO, SiO$_2$, O$_2$ etc., depending upon their dayside temperature and pressure \citep{2009ApJ...703L.113S, 2011ApJ...742L..19M, 2012ApJ...755...41S, 2021arXiv211204663W}. Since the surface of the planet can potentially be molten \citep{2011Icar..213....1L}, the composition of the atmosphere becomes strongly dependent on that of the interior \citep[see, e.g.,][]{2021arXiv211204663W}. The temperature gradient across the day- and night-side of the planet can transport some of the species to the night-side, causing their condensation \citep{2016ApJ...828...80K}. Additionally, some of the species can be lost into space \citep[see,][for a review]{2021arXiv211204663W}. These various processes could potentially result in temporal variability for the atmosphere of some USPs \citep{2016ApJ...828...80K}. In the most extreme cases, the surface of the planet itself is thought to be disintegrating and escaping into space, leading to the formation of dust tails that can be observed through their highly asymmetric transit light curves \citep[see, e.g.,][]{2012ApJ...752....1R, 2012A&A...545L...5B, 2014ApJ...784...40R}. The photometric and spectroscopic observations of such disintegrating planets, and more generally of USPs, can not only give insights into the nature of their atmospheres but also provide essential clues regarding their interior.


The formation mechanism of USPs is a topic of active research. Scarcity of close-in planets with sub-Jovian mass in the so-called Neptunian desert \citep[e.g.,][]{2007A&A...461.1185L, 2009MNRAS.396.1012D, 2011ApJ...727L..44S, 2013ApJ...763...12B, 2016NatCo...711201L}
which is thought to be a result of atmospheric escape \citep[e.g.,][]{2004A&A...418L...1L, 2012MNRAS.425.2931O, 2019AREPS..47...67O}, led to a conjecture that the USPs are remnants of hot Neptunes that have gone through strong mass loss \citep[e.g.,][]{2011A&A...529A.136E, 2013ApJ...776....2L, 2020A&A...644A.174V}. Furthermore, \citet{2017AJ....154...60W} show that the metallicity distribution of USP hosts is very similar to that of mini-Neptune hosts, which supports the suggested formation pathway. Other hypotheses proposed to explain the formation of USPs include the migration of rocky planets from outer orbits and in-situ formation \citep{2008MNRAS.384..663R, 2018NewAR..83...37W}.


In order to better understand the formation mechanism of USPs and to constrain their internal structure and atmospheric composition, it is important to perform systematic studies refining their mass and radius. TESS, with its whole sky survey mode for exoplanet discovery, is a unique instrument to discover new USPs around nearby stars. \citet{2021AJ....161...56W} and \citet{2021MNRAS.501.4148L} have uncovered a multiplanetary system around a small, metal-poor G-dwarf TOI-561, which contains a USP, TOI-561\,b, by analysing two of TESS sectors along with other follow-up observations. Being a kinematic and chemically thick disk star, TOI-561, is alpha-enhanced \citep{2022MNRAS.511.4551L}. This means that composition of the proto-planetary disk is likely different to those around Solar-metallicity stars. This chemical-diversity is likely propagated through to planetary internal structures \citep{Asplund2009,Thiabaud+2014}, as is starting to be seen observationally \citep{Adibekyan+2021,Chen2022,2022MNRAS.511.1043W}. This could makes the core of the planets it hosts smaller and silicate mantle larger. This, in addition to potentially volatile envelope, is likely one of the main reasons making TOI-561 b the lowest density USP observed up to today. The low density composition \citep[$0.69\,\rho_\oplus$,][]{2022MNRAS.511.4551L} is drastically different from other well-known USPs, such as 55 Cnc e or K2-141\,b, which have much higher densities \citep[$1.21\,\rho_\oplus$ and $1.48\,\rho_\oplus$, respectively,][]{2018A&A...619A...1B, 2018AJ....155..107M}. This makes TOI-561\,b an interesting target for follow-up observations.


Among the multiple planets hosted by TOI-561, the inner-most planet was found to orbit the star in just $\sim 11$ hours. To probe the nature of this extremely irradiated planet, which is also the lowest density USP observed until today, we observed 13 new transits of TOI-561\,b with the CHaracterising ExOPlanet Satellite \citep[CHEOPS;][see Section \ref{sec:obs_ana}]{2021ExA....51..109B} to refine its radius, and thus improve our knowledge of its internal structure and putative atmosphere. We combined the new CHEOPS observations along with three CHEOPS archival transits and four TESS sectors to update the planetary parameters (Section \ref{sec:obs_ana}). We also searched for an occultation signal in the available dataset to learn about the planetary atmosphere. Using the updated parameters, we performed internal structure modelling of the planet (Section \ref{sec:internal_structure}). Section \ref{sec:atmos} explores various theoretical models to understand the emission from the planet (Section \ref{subsec:pandexo}). Subsequently, we use these models to make predictions for observations with the James Webb Space Telescope (JWST).
We then carried out an analysis to search for variability in the transit depth (Section \ref{subsec:variability}). We discuss our conclusions and their implications for future work in Section \ref{sec:conclusions}.
\section{Observations and analysis}\label{sec:obs_ana}

\begin{table*}
    \centering
    \caption{Observation log for TESS and CHEOPS}
    \begin{tabular}{cccc}
    \hline
    \hline
    \noalign{\smallskip}
    Visit/Sector & Start Date & End Date & File Key     \\
    \noalign{\smallskip}
    \hline
    \hline
    \noalign{\smallskip}
    TESS Observations & & & \\ \hline
    8 & {\small 2019-02-02 20:49:29} & {\small 2019-02-27 12:07:39} & Sector 8 \\
    35 & {\small 2021-02-09 11:51:16} & {\small 2021-03-06 11:36:59} & Sector 35 \\
    45 & {\small 2021-11-07 11:45:08} & {\small 2021-12-02 03:00:39} & Sector 45 \\
    46 & {\small 2021-12-03 01:34:47} & {\small 2021-12-30 04:54:14} & Sector 46 \\\hline
    \noalign{\smallskip}
    CHEOPS Observations & & & \\ \hline
    \noalign{\smallskip}
    37001 & {\small 2021-01-23 15:29:08} & {\small 2021-01-24 07:09:34} & {\small CH\_PR100031\_TG037001} \\
    801 & {\small 2021-02-28 20:16:29} & {\small 2021-03-01 00:41:36} & {\small CH\_PR100008\_TG000801}\\
    101 & {\small 2021-03-01 01:12:29} & {\small 2021-03-01 13:56:50} & {\small CH\_PR110050\_TG000101}\\
    802 & {\small 2021-03-06 04:53:31} & {\small 2021-03-06 10:18:40} & {\small CH\_PR100008\_TG000802}\\
    803 & {\small 2021-03-08 10:30:28} & {\small 2021-03-08 15:19:36} & {\small CH\_PR100008\_TG000803}\\
    804 & {\small 2021-03-12 11:02:08} & {\small 2021-03-12 15:26:15} & {\small CH\_PR100008\_TG000804}\\
    805 & {\small 2021-03-16 00:55:08} & {\small 2021-03-16 05:44:16} & {\small CH\_PR100008\_TG000805}\\
    806 & {\small 2021-03-20 02:04:08} & {\small 2021-03-20 06:28:15} & {\small CH\_PR100008\_TG000806}\\
    807 & {\small 2021-03-20 22:11:48} & {\small 2021-03-21 03:21:57} & {\small CH\_PR100008\_TG000807}\\
    808 & {\small 2021-03-23 03:48:32} & {\small 2021-03-23 08:36:40} & {\small CH\_PR100008\_TG000808}\\
    809 & {\small 2021-03-24 02:09:09} & {\small 2021-03-24 06:58:17} & {\small CH\_PR100008\_TG000809}\\
    810 & {\small 2021-03-25 00:00:07} & {\small 2021-03-25 04:42:15} & {\small CH\_PR100008\_TG000810}\\
    811 & {\small 2021-03-29 10:19:08} & {\small 2021-03-29 14:44:15} & {\small CH\_PR100008\_TG000811}\\
    812 & {\small 2021-03-30 08:23:29} & {\small 2021-03-30 13:12:37} & {\small CH\_PR100008\_TG000812}\\
    813 & {\small 2021-03-31 06:01:29} & {\small 2021-03-31 10:50:37} & {\small CH\_PR100008\_TG000813}\\
    39301 & {\small 2021-04-12 23:52:28} & {\small 2021-04-15 05:37:57} & {\small CH\_PR100031\_TG039301} \\
    \noalign{\smallskip}
  \hline
    \end{tabular}
    \tablefoot{The file key is a unique identifier of each of the CHEOPS observations, and helps in retrieving data from the archive.}
    \label{tab:obs_log}
\end{table*}



\subsection{Observations and data reduction}\label{subsec:data}

We observed precise transit photometry of TOI-561 with CHEOPS, which is an S-class mission launched by the European Space Agency to obtain precision photometry of exoplanetary systems. We observed 13 transits of TOI-561\,b with CHEOPS within the Guaranteed Time Observation (GTO) program during February-March 2021 (see, Table \ref{tab:obs_log} for the observation log). In addition to this, we included in our analysis three archival visits of CHEOPS for the same system from \citet{2022MNRAS.511.4551L}. Since the target is moderately bright ($m_{\textnormal{Gaia}} = 10.01$), we observed it with the longest available exposure time of 60 s to reduce the instrumental noise. To save bandwidth, only the so-called sub-arrays, a cropped circular subsections of 100 pixel radius, are downloaded instead of the full frame images. The photometry from these sub-arrays was extracted with the CHEOPS Data Reduction Pipeline \citep[DRP;][]{2020A&A...635A..24H}, which uses aperture photometry to derive fluxes from the sub-arrays. We also extracted photometry independently using the PSF photometry package \texttt{PIPE}\footnote{\url{https://github.com/alphapsa/PIPE}} (see \citealt{2021A&A...654A.159S} for more information). When comparing the extracted lightcurves, we found \texttt{PIPE} to typically produce a $\sim$10\% lower scatter at around 300 part per million (ppm) as measured by the median absolute deviation (MAD) at the observed time cadence of 1\,min. In the following we therefore report on the analysis of the PIPE photometry, although the results are consistent with independent analysis of the DRP photometry that was also made but that is not reported here.

Prior to performing the transit analysis, we discarded all points from our CHEOPS dataset which are flagged by \texttt{PIPE} to have poor photometry. This is mostly because those points are contaminated by, e.g., strong cosmic rays or trails from satellites crossing the field. Additionally, we discarded frames with high background (higher than 400\,e$^-$\,pix$^{-1}$, mostly when pointing near Earth's limb) as the background often shows a temporally variable spatial structure that is difficult to remove and significantly increases the noise. The photometry also shows a trend with background at high values, possibly due to a non-linearity in the charge transfer efficiency. This filtering of data points is independent of the photometric values, so does not bias the light curve. About 12\% of data points are discarded in this way, almost all from observations with a line of sight within 20$^{\circ}$ from the Earth limb.

In addition to the CHEOPS datasets we used TESS data from the four sectors in which it observed TOI-561 (Sectors 8, 35, 45 and 46). We used the PDC-SAP photometry \citep{2012PASP..124.1000S, 2014PASP..126..100S} reduced by the TESS Science Processing Operations Center \citep[SPOC;][]{2016SPIE.9913E..3EJ}.

\subsection{Transit lightcurve analysis}\label{subsec:transit_ana}
Although we take advantage of both the CHEOPS and TESS datasets to refine the planetary properties of TOI-561\,b, we first analysed these datasets individually. This was mainly to determine and constrain the astrophysical and systematic noise models for CHEOPS and TESS data. We use \texttt{juliet} \citep{2019MNRAS.490.2262E} to perform this analyses. We detail this process in the sub-sections below.

TOI-561\,b is the innermost body in the four-planet system around TOI-561. Since we focus on TOI-561\,b in the present work we decided to set wide uninformative priors for its planetary parameters, except for the orbital period and the transit time on which we set Gaussian priors based on \citet{2022MNRAS.511.4551L}. Since TOI-561\,b orbits the star in around 11 hours, we expect its orbit to be circularised. Accordingly we fixed its orbital eccentricity ($e_b$) to zero and argument of periastron ($\omega_b$) to $90^\circ$. For the other three planets we set Gaussian priors on most of the planetary parameters, with mean and standard deviation based on their values from \citet{2022MNRAS.511.4551L}. We fixed their eccentricities and arguments of periastron passage to the values from \citet{2022MNRAS.511.4551L}, as it is difficult to retrieve these properties from the photometric data alone. The values from \cite{2022MNRAS.511.4551L} are expected to be robust as they use plenty of radial velocity (RV) data points along with photometric data to estimate those parameters. The latter is also the reason, in addition to the lack of further public RV data, we decided not to use RV data in our analysis.

Instead of using different priors for the scaled semi-major axis ($a/R_\star$) of each planet, we fitted the stellar density to take advantage of our knowledge of stellar properties. The value of the stellar density, which can be connected to the scaled semi-major axis through Kepler's third law, was computed from the stellar mass and radius reported in \citet{2022MNRAS.511.4551L}. Doing this we account for stellar properties while modelling planetary transits, reduce the number of free parameters in the fit by three, and enforce that all planets are orbiting a star with the same stellar density. For the sake of completeness, we re-did the whole analysis without using the stellar density prior (see, Appendix \ref{app:a}). The results are in agreement with our main approach albeit with increased uncertainties, as expected.

We use quadratic limb darkening law in our analysis. We parametrize the limb darkening coefficients (LDCs) as suggested by \citet{2013MNRAS.435.2152K} and use this parametrization in our model. This was to ensure that they yield physically plausible stellar intensity profile. As the theoretical LDCs, computed from model stellar atmospheres, could have discrepancies with the empirical LDCs \citep[see, e.g.,][]{2015MNRAS.450.1879E, 2022AJ....163..228P}, we set uniform priors between 0 and 1 to the transformed quadratic LDCs, for both CHEOPS and TESS bandpasses. The full list of priors on the planetary parameters can be found in Table \ref{tab:fitted_transit_param}.

\subsubsection{Analysis of the CHEOPS photometry}\label{subsubsec:cheops_ana}
The orbit of CHEOPS is designed to be nadir locked, with the satellite revolving continuously over the day-night terminator of the Earth. As a result, the field-of-view of CHEOPS rotates around the target star in the image frame. The rotation can introduce correlations with background stars. These environmental effects, in addition to other instrumental effects, makes CHEOPS photometry correlated with instrumental parameters \citep{2020A&A...643A..94L, 2022MNRAS.511.1043W}. The transit lightcurves that we obtained for TOI-561\,b were no exception ---  we therefore detrend the data against spacecraft roll angle in order to provide robust estimates of the planetary properties.

In most of the visits we analysed, the trend with roll angle was too complicated to model with sinusoidal functions. Thus, a Gaussian Process (GP) model, implemented from \texttt{celerite} \citep{2017AJ....154..220F} in \texttt{juliet}, built from an Exponential-Mat\'{e}rn kernel, was used to model the trend with roll angle. Each visit was decorrelated individually against the roll angle in this way. However, as the number of data points in an individual visit is not enough to constrain the GP model properly, we fitted all visits simultaneously to obtain a well-constrained GP model that accounts for the trends with roll angle. In this whole dataset fit, the priors on GP hyperparameters were chosen from our previous analysis of individual visits to help convergence and a better optimization of parameter space. In the end, we subtracted the modelled trend with roll angle from our dataset to produce roll angle detrended photometry. To prevent an underestimation of uncertainties in the subsequent analysis, we propagated the uncertainties in the GP model by adding them in quadrature to the errorbars on fluxes of each visit \citep[similar to the treatment of CHEOPS data with PSF-basis vectors in][]{2022MNRAS.511.1043W}.

While the resulting photometry is corrected for the roll angle trend, it still shows the correlations with other parameters, such as the PSF centroid position, background, time etc. It is important to choose a suitable set of detrending vectors for each visit, otherwise we may overfit or underfit the dataset. Our systematic search for a set of decorrelation vectors was done using \texttt{pycheops} \citep{2021MNRAS.tmp.3057M}. In this search we added one by one decorrelation vector to our model and kept the vector only if it yielded a higher Bayes' factor. The optimal sets of detrending vectors obtained in this way contain background and up to second order polynomials in PSF centroid position. In our analysis, we use a linear model to detrend against these vectors.

Some of the visits also show the well understood ``ramp'' effect in the dataset \citep{2021A&A...653A.173M}. The ramp effect happens because of the change in the shape of PSF related to the thermal effects in the telescope tube that arises mainly when re-pointing the satellite. This has been seen in other datasets and PSF-based methods have been seen to remove such trends \citep{2022MNRAS.511.1043W}. Since this effect is directly linked to the shape of the PSF, we can correct for it using the principal components of the PSF model extracted from \texttt{PIPE}. We added these components in our linear model when required.

In addition to these linear models, we also added a GP model (produced using Exponential-Mat\'{e}rn kernel from \texttt{celerite} implemented in \texttt{juliet}) to account for temporal astrophysical/systematic trends, meaning that our final model includes linear and GP models for a decorrelation and a transit model \citep[\texttt{batman};][]{2015PASP..127.1161K}. As previously, we first analysed each visit independently and then used the derived hyperparameters as priors in our joint photometry analysis.

\subsubsection{Analysis of the TESS photometry}\label{subsubsec:tess_ana}
We analysed data from the four TESS sectors. The main source of noise in TESS data is found to be systematic and astrophysical trends. To account for these trends we introduce in our fitting procedure a GP model based on an Exponential-Mat\'{e}rn kernel. Furthermore, our global model includes a four-planet transit model \citep[\texttt{batman};][]{2015PASP..127.1161K}, a jitter term and an out-of-transit flux offset. Since the PDC-SAP flux is expected to correct for the dilution from the nearby light sources, we set the dilution factor \citep[see][for details]{2019MNRAS.490.2262E} to 1 in our model, meaning that no dilution was assumed from nearby sources.

Similar to our CHEOPS analysis, we analysed individual TESS sectors before performing the joint analysis. The priors on the noise model hyperparameters in our joint analysis were selected based on the posteriors from the analysis of individual sectors.

\begin{table*}
    \caption{Planetary and stellar parameters used in the transit analysis}
    \centering
    \begin{tabular}{lcccr}
    \toprule
    \toprule
    Parameters & Symbols & Values & Priors & Units \\
    \midrule
    Planetary parameters (Planet b) &&&&\\
    \quad Orbital period & $P$ & {\small $0.4465689683 ^{+0.0000002381} _{-0.0000003152}$} & {\small$ \mathcal{N}(0.4465688, 7.5e-7)$} & days \\
    \noalign{\smallskip}
    \quad Transit time & $T_0$ & {\small $2459578.546253 ^{+0.000219} _{-0.000205}$} & {\small $\mathcal{N}(2459578.545979,0.000499)$} & BJD\\
    \noalign{\smallskip}
    \quad Planet-to-star radius ratio & $R_p/R_\star$ & {\small $0.015439 ^{+0.000218} _{-0.000226}$} & {\small $\mathcal{U}(0,1)$} & -\\
    \noalign{\smallskip}
    \quad Impact parameter & $b$ & {\small $0.088469 ^{+0.050057} _{-0.047097}$} & {\small $\mathcal{U} (0,1)$} & - \\
    \noalign{\smallskip}
    \quad Scaled semi-major axis & $a/R_\star$ & {\small $2.692202 ^{+0.017130} _{-0.021185}$} & - & -\\
    \noalign{\smallskip}
    \quad Eccentricity & $e_b$ & - & Fixed to 0 & -\\
    \noalign{\smallskip}
    \quad Argument of periastron passage & $\omega_b$ & - & Fixed to 90 & deg\\
    \midrule
    Stellar Parameters &&&& \\
    \quad Stellar density & $\rho_\star$ & {\small $1850.964990 ^{+35.557225} _{-43.352182}$} & {\small $\mathcal{N}(1896.89, 91.17)$} & kg/m$^3$ \\
    \noalign{\smallskip}
    \quad Limb darkening coefficients &&&& \\
    \multirow{2}{*}{\qquad CHEOPS passband} & $u_{1_{CHEOPS}}$ & {\small $0.291191 ^{+0.150360} _{-0.149110}$} & - & - \\
    \noalign{\smallskip}
    & $u_{2_{CHEOPS}}$ & {\small $0.397271 ^{+0.242151} _{-0.238942}$} & - & - \\
    \noalign{\smallskip}
    \multirow{2}{*}{\qquad TESS passband} & $u_{1_{TESS}}$ & {\small $0.357040 ^{+0.146761} _{-0.170364}$} & - & - \\
    \noalign{\smallskip}
    & $u_{2_{TESS}}$ & {\small $0.226039 ^{+0.259467} _{-0.232569}$} & - & - \\
    \noalign{\smallskip}
    \midrule
    Derived planetary Parameters\tablefootmark{$\dagger$} &&&& \\
    \quad Radius & $R_p$ & {\small $1.4195 ^{+0.0217} _{-0.0224}$} & - & $R_\oplus$ \\
    \noalign{\smallskip}
    \multirow{2}{*}{\quad Density} & $\rho_p$ & {\small $4.3049 ^{+0.4411} _{-0.4216}$} & - & g/cm$^3$ \\
    \noalign{\smallskip}
    & $\rho_p$ & {\small $0.7834 ^{+0.0803} _{-0.0767}$} & - & $\rho_\oplus$ \\
    \noalign{\smallskip}
    \quad Orbital distance & $a_b$ & {\small $0.0106 ^{+0.0001} _{-0.0001}$} & - & AU \\
    \noalign{\smallskip}
    \quad Inclination & $i_b$ & {\small $88.1178 ^{+1.0045} _{-1.0820}$} & - & deg \\
    \bottomrule
    \end{tabular}
    \tablefoot{The Gaussian priors with mean $\mu$ and variance $\sigma^2$ are displayed as $\mathcal{N}(\mu, \sigma^2)$. $\mathcal{U}(a,b)$ shows the uniform prior between $a$ and $b$.\\ \tablefoottext{$\dagger$}{In the calculation of the derived planetary parameters we used following stellar and planetary parameters from \citet{2022MNRAS.511.4551L} and \citet{Brinkman2023}: {\small $R_\star=0.843 \pm 0.005\,R_\odot$} and {\small $M_p = 2.24 \pm 0.20 \, M_\oplus$}.}}
    \label{tab:fitted_transit_param}
\end{table*}

\subsubsection{Joint photometry analysis}\label{subsubsec:joint_ana}
Our final analysis consists in a joint fit to the TESS and CHEOPS datasets, in order to refine the planetary parameters for TOI-561\,b as much as possible. As previously, our global model contains, in addition to a four-planets transit model, a jitter term, a mean out-of-transit offset and linear and GP models to account for various systematic and astrophysical correlations in the data. We used our earlier analysis of individual CHEOPS visits and TESS sectors to set informative priors on the nuisance parameters in the joint CHEOPS-TESS analysis. Adapting this two-step method not only helps in determining and constraining the noise model, but also allows for a better and faster sampling of the parameter space. The second point is crucial when the total number of free parameters becomes large, which is the case in the present analysis.

Indeed, the four-planets transit model and different noise models for each visit/sector yields a large number of free parameters in the analysis, for a grand total of 168. Among these, 143 are nuisance parameters accounting for various decorrelations in the datasets. The other parameters are either planetary or stellar. We decided to use nested sampling methods \citep{2004AIPC..735..395S,10.1214/06-BA127} to sample the distribution from the posterior. Since we are dealing with a high dimensional parameter space, we follow the recommendations from \citet{2019MNRAS.490.2262E} and use dynamic nested sampling \citep{2019S&C....29..891H} as included in \texttt{juliet} via \texttt{dynesty} \citep{2020MNRAS.493.3132S}.

\begin{figure}
    \centering
    \includegraphics[width=\columnwidth]{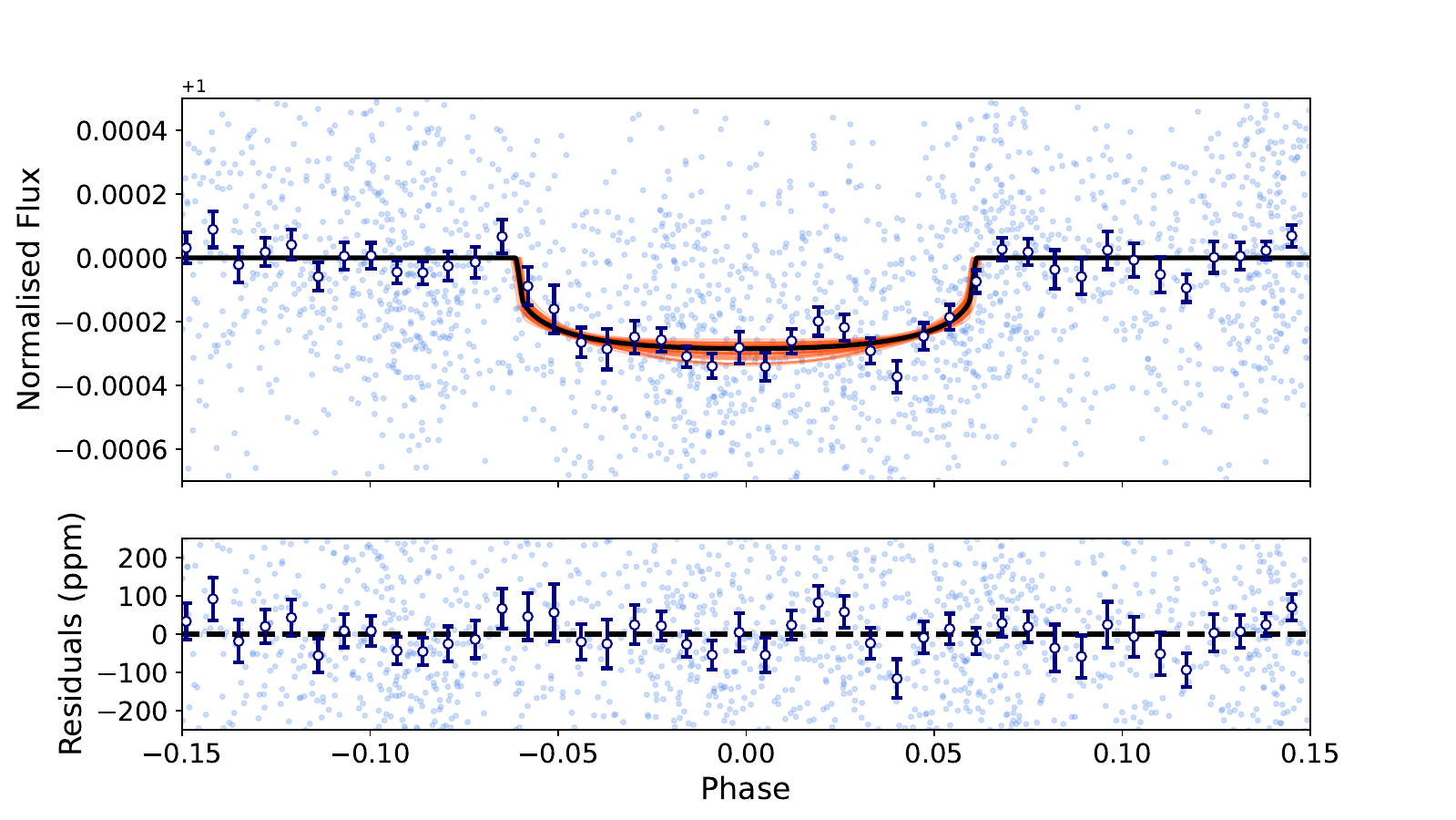}
    \includegraphics[width=\columnwidth]{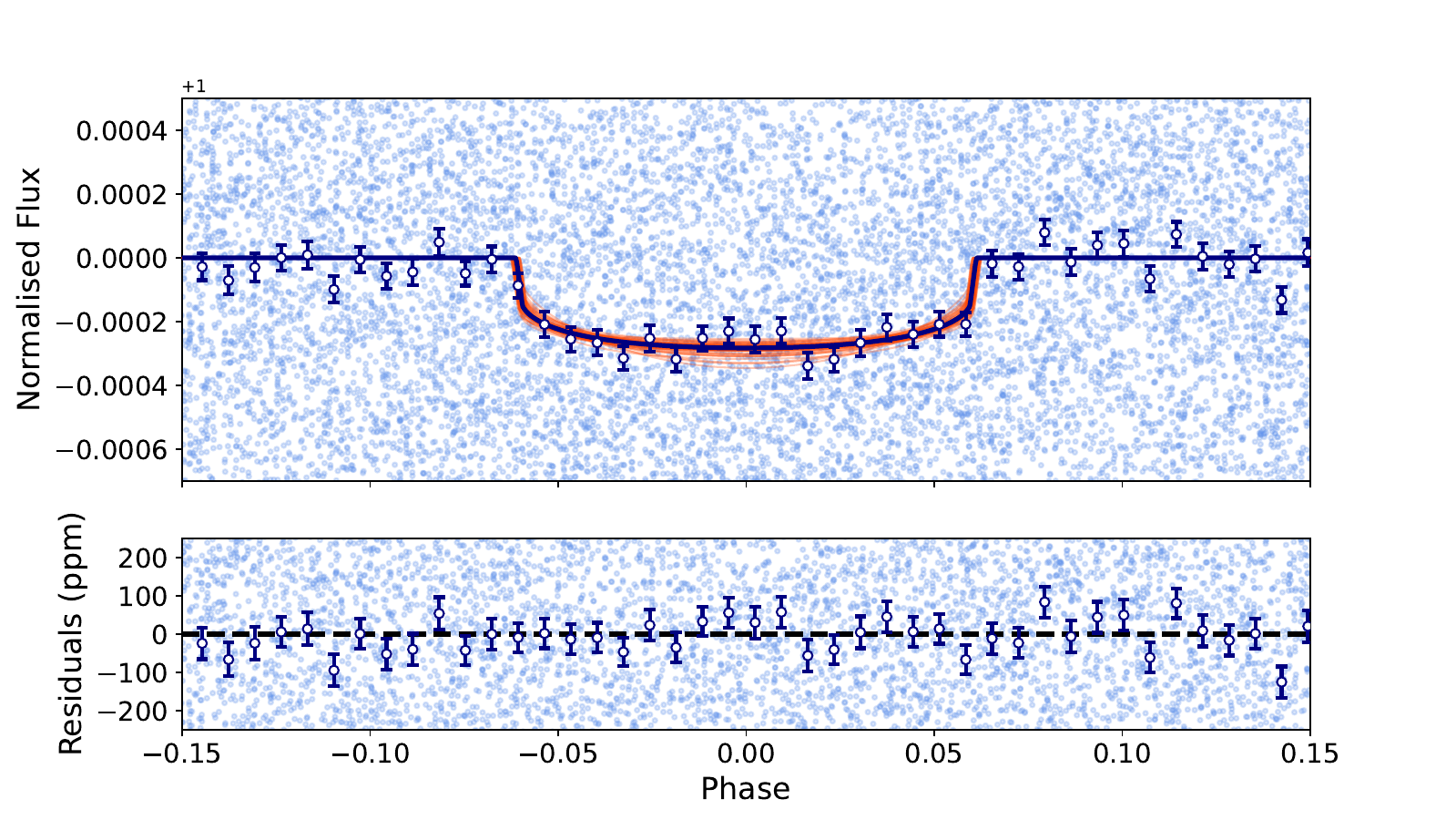}
    \caption{Phase-folded lightcurve for planet b, over all of the \textit{(a)} CHEOPS visits and \textit{(b)} TESS sectors. \textit{Top sub-panels}: the dark blue curve represents the median fitted model, and the orange curves are models computed from randomly drawn samples from the posterior. The light blue and dark blue points are the original points and the binned data points. \textit{Bottoms sub-panels} residuals from the median model.}
    \label{fig:phase_folded}
\end{figure}

The fitted median transit model for planet b, along with detrended data and residuals from the fit, are shown in Figure \ref{fig:phase_folded} for CHEOPS and TESS. It can be seen from the residuals that all of the instrumental and astrophysical trends were effectively removed, demonstrating the quality of our fitting procedure. The corresponding median posteriors and their 1-$\sigma$ credible intervals for various planetary parameters are tabulated in Table \ref{tab:fitted_transit_param}. Raw and detrended data, along with the best-fitted models, for individual CHEOPS visits are displayed in Figure \ref{fig:all_photometry} in the Appendix \ref{app:b}. The correlation plot, made using \texttt{corner.py} \citep{2016JOSS....1...24F}, for fitted planetary parameters is presented in Figure \ref{fig:corner_plot} in the Appendix \ref{app:c}.

\begin{figure}
    \centering
    \includegraphics[width=\columnwidth]{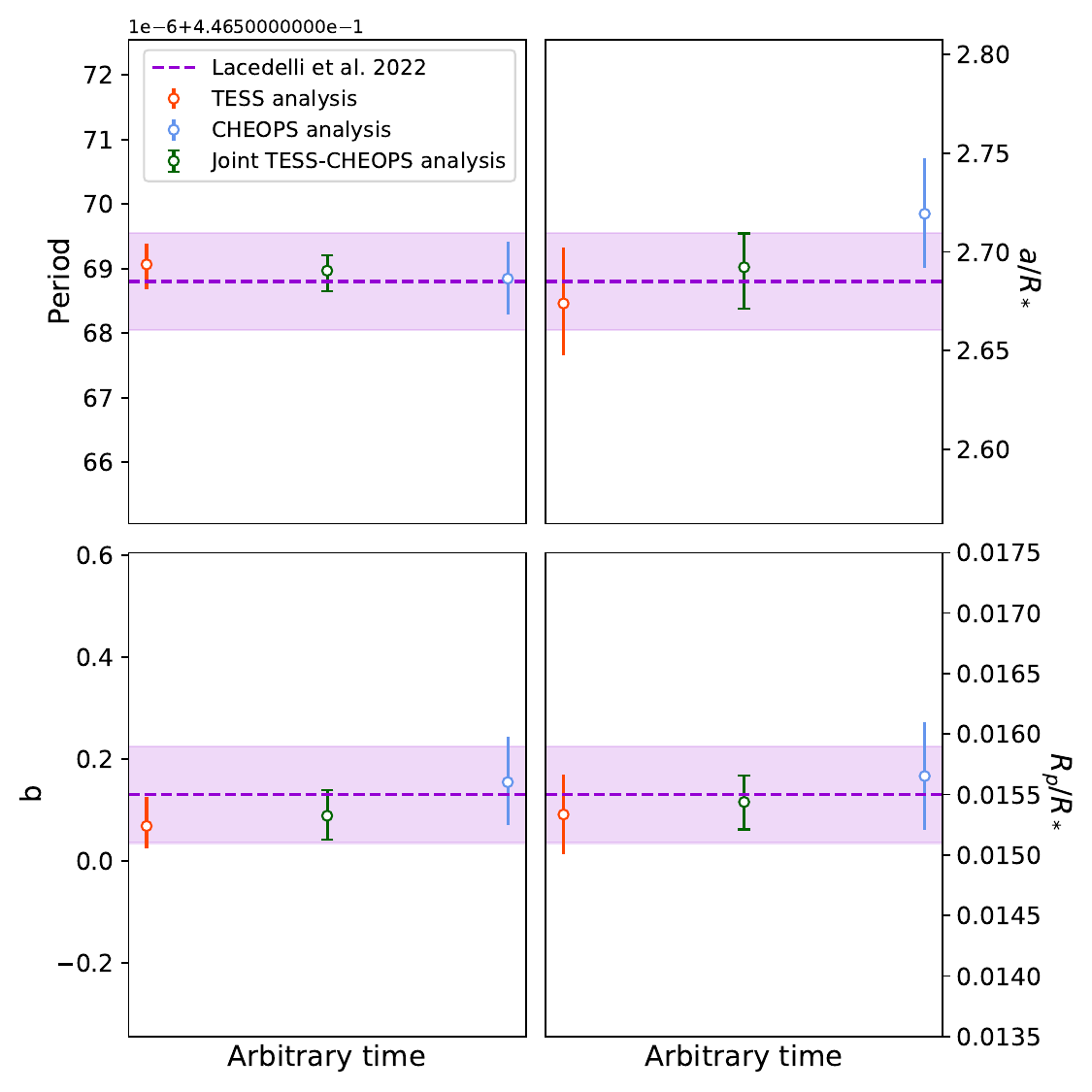}
    \caption{Comparison of retrieved planetary parameters ($P$, $a/R_\star$, $b$ and $R_p/R_\star$) from our analyses of CHEOPS (blue points), TESS (red points) and joint dataset (green points) with the literature values from \citet{2022MNRAS.511.4551L}. The latter is represented by the dashed purple line and $1-\sigma$ uncertainty band. The horizontal axis gives the arbitrary transit time. Note that the scale on the vertical axis is relative and centered on the literature value of the given parameter.}
    \label{fig:plan_param_comp}
\end{figure}

In Figure \ref{fig:plan_param_comp} we show an inter-comparison of $P$, $a/R_\star$, $b$ and $R_p/R_\star$ between our various analysis, and with their literature values. It can readily be observed that our CHEOPS and TESS analysis agree with each other and with the joint analysis, and that they are all in excellent agreement with the values reported from \citet{2022MNRAS.511.4551L}. Thanks to the additional CHEOPS and TESS data we are able to put stronger constraints on the planetary parameters, in particular the planet-to-star radius ratio with a precision improved to $\sim 2\%$.

\subsection{Eclipse analysis}\label{subsec:eclipse}

\begin{figure}
    \centering
    \includegraphics[width=\columnwidth]{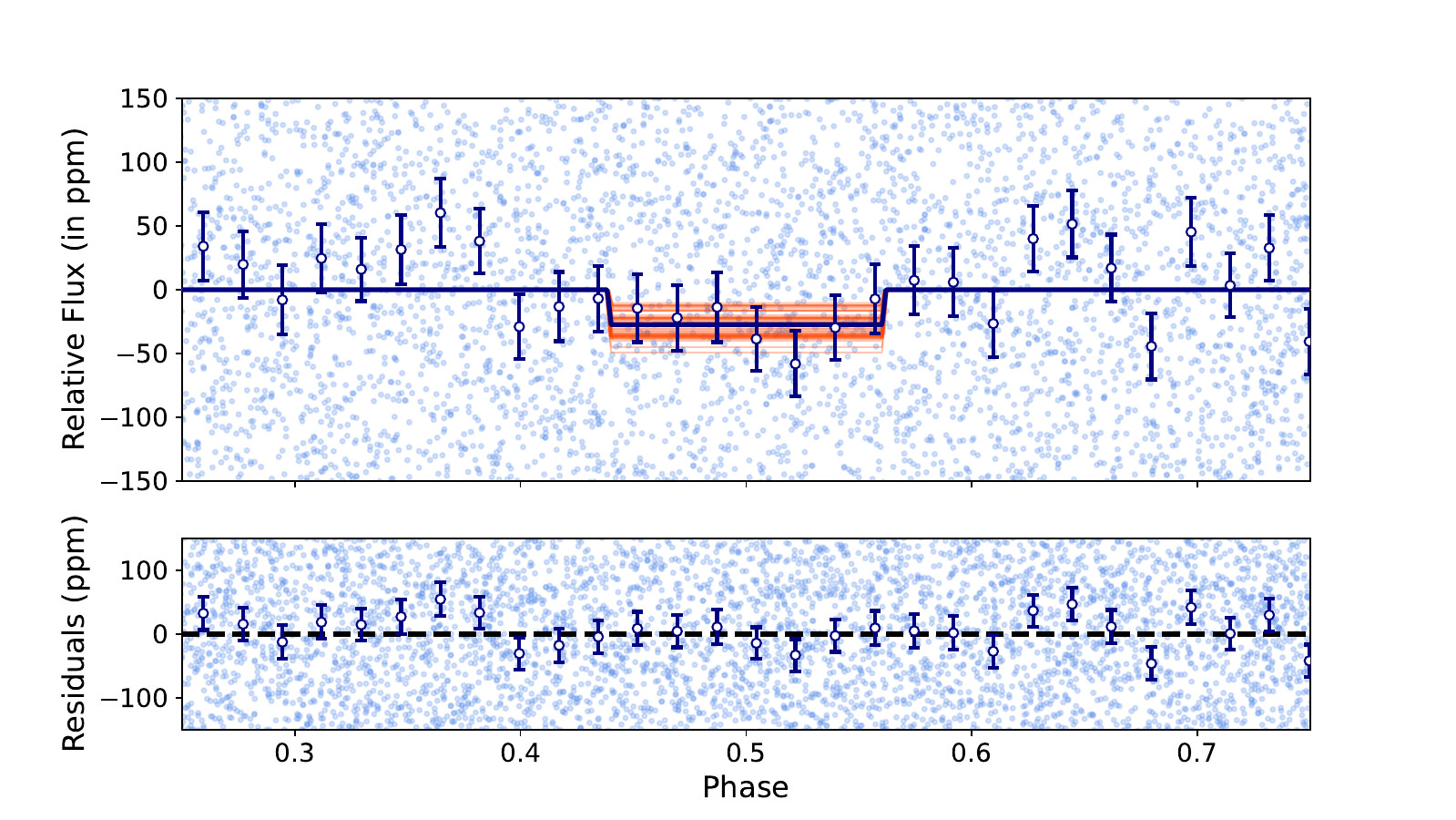}
    \includegraphics[width=\columnwidth]{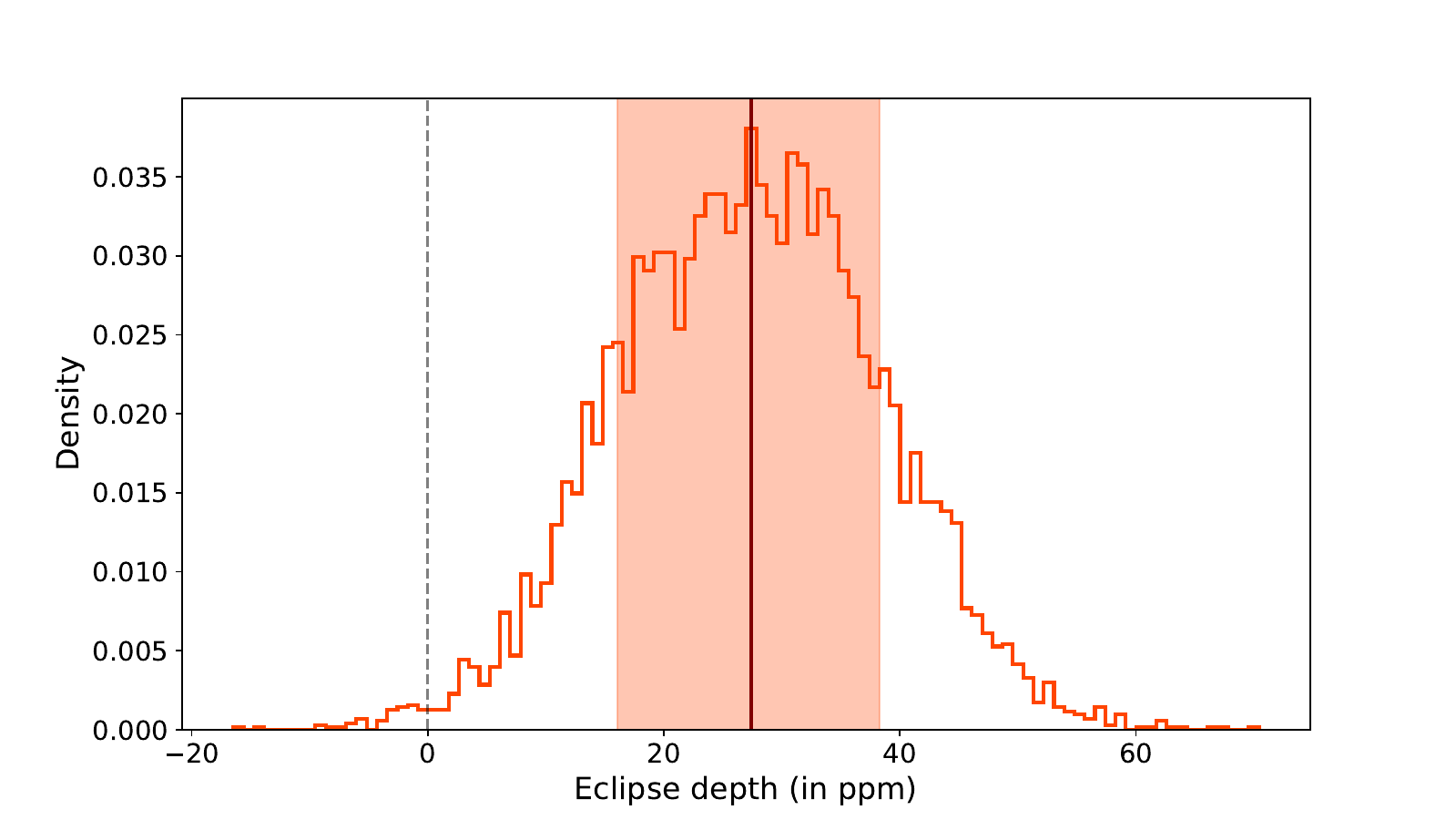}
    \caption{\textit{Top panel}: The upper plot shows the median fitted model (in dark blue), which includes the planet transit and eclipse,  and the original (in light blue) and binned datapoints (dark blue) from four TESS sectors. The orange curves are the models computed from the randomly chosen samples from the posteriors. The lower plot shows the residuals for the fit. \textit{Bottom panel}: Posterior distribution of the eclipse depth. The dark orange line and the light orange band gives the median eclipse depth and the $1-\sigma$ credible interval. The dashed line represents the null hypothesis.}
    \label{fig:phase_folded_eclipse}
\end{figure}

TOI-561\,b orbits its host star at a short orbital distance of $0.0105$\,AU (see, Table \ref{tab:fitted_transit_param}). We expect the dayside temperature\footnote{Calculated using, $T_{\textnormal{day}} = T_\star \sqrt{\frac{R_\star}{a}} (1-A_B)^{1/4} f^{1/4}$, while using the bond albedo $A_B=0$ and the heat re-distribution factor $f=2/3$, for a bare rock with no heat re-distribution and $f=1/4$ for uniform heat distribution \citep{2014PNAS..11112601B, 2019ApJ...886..140K}.} of this planet to be around {\small $T_{\textnormal{day}} \sim 2963.51$\,K} for instant re-radiation of the heat to the space (i.e., no heat re-distribution) and {\small $T_{\textnormal{day}} \sim 2319.07$\,K} for a uniform heat distribution \citep[effective temperature of the host star {\small $T_\star = 5372$\,K};][]{2022MNRAS.511.4551L}. Based on these estimates we foresee the planet to produce strong thermal emission, even in the optical CHEOPS passband. In the light of previous CHEOPS measurements of planetary occultations \citep[e.g. ][]{2020A&A...643A..94L, 2022A&A...658A..75H}, we thus search for a secondary eclipse signal in our dataset.

As our CHEOPS observations are focused on the transit event, most of the exposures do not cover the secondary eclipse of the planet. Fortunately though, two of the archival visits (Visit No. 39301 and 101, see, Table \ref{tab:obs_log}) cover orbital phases during which occultation occurs  five times. However we found that these five occurences are not enough to recover any significant signal from the CHEOPS data. Nonetheless, we derive a 99.7 percentile limit of 98.93 ppm on eclipse depth for this dataset.

On the other hand, TESS observed the event more than 150 times during its four observation sectors. Furthermore, we anticipate a larger occultation signal in the TESS data considering that the thermal contribution from TOI-561\,b would be larger in the TESS bandpass than in the CHEOPS bandpass. With our upper estimation for the planet temperature {\small $T_{\textnormal{day}} \sim 2963.51$\,K} we expect the magnitude of the occultation to be at least $14$ ppm in the TESS bandpass assuming blackbody emission. To test this hypothesis, we modelled the TESS data with a joint transit-eclipse model and an only-transit model using \texttt{batman} models implemented in \texttt{juliet}. We masked out transit signals from all other planets to simplify the analysis. We use the same method as described in Section \ref{subsec:transit_ana} to model the dataset, but this time we provide informative priors on all planetary parameters except the eclipse depth. We set a uniform prior between $-100$\,ppm to $100$\,ppm for the eclipse depth. We compare the Bayesian evidence for a model comparison between the model with and without eclipse. 

We find that the model with eclipse is strongly favoured statistically ($\Delta \ln{Z} \sim 1.43$). The phase-folded light curve, including the median model with randomly selected models, is shown in Figure \ref{fig:phase_folded_eclipse}. We are only able to put an upper limit on the eclipse depth (a 99.7 percentile limit of $60.15$\,ppm) because the value we derive from its posterior distribution (Figure \ref{fig:phase_folded_eclipse}) is consistent with zero at 3-$\sigma$: {\small $L=27.40 ^{+10.87} _{-11.35}$\,ppm}. The derived posterior can still put some constraints on the planetary atmosphere.

\begin{figure}
    \centering
    \includegraphics[width=\columnwidth]{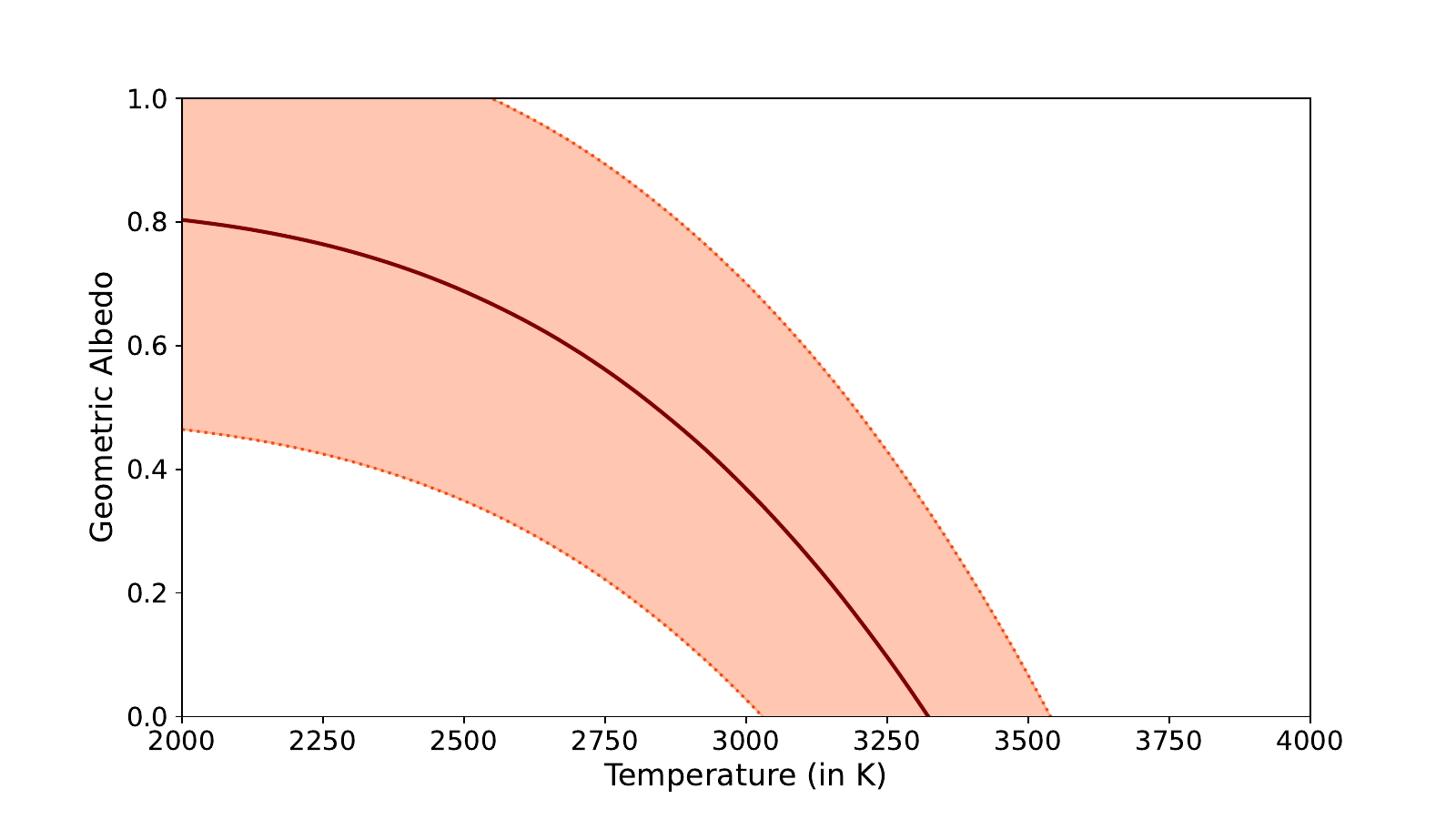}
    \caption{Relation between the geometric albedo and the brightness temperature of the planet, given the posteriors of the occultation depth (with mean and $1-\sigma$ intervals $27.40 ^{+10.87} _{-11.35}$\,ppm) in the TESS bandpass.}
    \label{fig:ag_vs_temp}
\end{figure}

It should be noted here that both reflection and thermal emission could contribute to the occultation signal we are detecting in the TESS data. With TESS data alone we cannot determine the fractional contribution to the eclipse depth from these components, so that we would need similar observations at longer wavelengths. With the measured posterior of eclipse depth from the TESS data we can, however, estimate the geometric albedo ($A_g$) over a range of brightness temperatures (a measure of the thermal emission) for this planet. We followed the method from \citet{2019A&A...624A..62M} to find the relation between the geometric albedo and the temperature, which is plotted in Figure \ref{fig:ag_vs_temp}. As expected, the median solution (depicted by an orange line) is degenerate. The posterior of eclipse depth (as shown in Figure \ref{fig:phase_folded_eclipse}, with $L = 27.40 ^{+10.87} _{-11.35}$\,ppm), assuming a zero contribution from reflection, would result in a dayside temperature of $\sim 3325$\,K. On the other hand, to explain the extracted eclipse depth with reflection alone, the geometric albedo of the planet should be around $0.83$. Note that both of these values are the extreme cases and the true solution should lie somewhere in between.

\section{Internal structure modelling}\label{sec:internal_structure}

\begin{table*}
    \caption{Posteriors of the internal structure parameters}
    \label{tab:int_struc_water}
    \centering
    \begin{tabular}{lrr}
    \toprule
    \toprule
    Internal structure parameter & Model with water layer & Model without water layer \\
    \midrule
    M\textsubscript{core}/M\textsubscript{total} & {\small $0.09^{+0.10}_{-0.08}$} & {\small $0.12^{+0.11}_{-0.11}$} \\
    \noalign{\smallskip}
    M\textsubscript{water}/M\textsubscript{total} & {\small $0.22^{+0.16}_{-0.15}$} & - \\
    \noalign{\smallskip}
    log M\textsubscript{gas} [M\textsubscript{$\oplus$}] & {\small $-9.48^{+2.20}_{-2.27}$} & {\small $-7.06^{+0.24}_{-0.28}$} \\
    \noalign{\smallskip}
    \bottomrule
    \end{tabular}
    \tablefoot{Posteriors parameters for models with and without an included fully distinct water layer. The errors are the 5 and 95 percentile of the corresponding posterior distributions.}
\end{table*}

\begin{table*}
    \caption{Posteriors of the internal structure parameters for models with fixed (to stellar values) and varying Si/MG/Fe ratios in the models}
    \label{tab:int_struc_comp}
    \centering
    \begin{tabular}{lrr}
    \toprule
    \toprule
    Internal structure parameter & Model with fixed Si/Mg/Fe ratios & Model with free Si/Mg/Fe ratios \\
    \midrule
    M\textsubscript{core}/M\textsubscript{total} & {\small $0.09^{+0.10}_{-0.08}$} & {\small $0.06^{+0.14}_{-0.05}$} \\
    \noalign{\smallskip}
    M\textsubscript{water}/M\textsubscript{total} & {\small $0.22^{+0.16}_{-0.15}$} & {\small $0.27^{+0.18}_{-0.20}$} \\
    \noalign{\smallskip}
    log M\textsubscript{gas} [M\textsubscript{$\oplus$}] & {\small $-9.48^{+2.20}_{-2.27}$} & {\small $-9.40^{+2.23}_{-2.34}$} \\
    \noalign{\smallskip}
    Fe\textsubscript{core} & {\small $0.90^{+0.09}_{-0.08}$} & {\small $0.90^{+0.09}_{-0.08}$} \\
    \noalign{\smallskip}
    Si\textsubscript{mantle} & {\small $0.41^{+0.07}_{-0.06}$} & {\small $0.42^{+0.08}_{-0.07}$} \\
    \noalign{\smallskip}
    Mg\textsubscript{mantle} & {\small $0.48^{+0.10}_{-0.10}$} & {\small $0.49^{+0.12}_{-0.23}$} \\
    \noalign{\smallskip}
    \bottomrule
    \end{tabular}
    \tablefoot{Posteriors for two models, the first model assumes the Si/Mg/Fe ratios of the planet match the ones measured for the star exactly and the other model where a wide variety of Si/Mg/Fe ratios are allowed. The errors are the 5 and 95 percentile of the corresponding posterior distributions.}
\end{table*}

In the following, we discuss the internal structure of TOI-561\,b. The method used is described in more depth in \citet{Leleu+2021} and is based on \citet{Dorn+2017}, but we will briefly summarise the most important aspects below. The Bayesian inference scheme we applied takes as input parameters the stellar (mass, radius, effective temperature, age and [Si/H], [Mg/H] and [Fe/H]) and planetary observables (mass relative to the star, transit depth and period) of the system. The likelihood of a given structure is calculated based on an internal structure model. We assume that the planet is spherically symmetric and consists of four fully distinct layers: an inner iron core \citep{Hakim+2018}, a silicate mantle \citep{Sotin+2007}, a water layer \citep{Haldemann+2020} and a pure H/He atmosphere \citep{Lopez+Fortney2014}. Furthermore, we assume that the Si/Mg/Fe ratios of the planet match the ones of the star exactly. Although this is supported e.g. by \citet{Thiabaud+2015}, recent work by \citet{Adibekyan+2021} suggests that the correlation might not be 1:1. Implementing this possibility in the model is the subject of future work.

For the priors of the internal structure parameters, we chose a prior that is uniform in log for the gas mass. For the mass fractions of the inner core, mantle and water layers with respect to the solid planet, our chosen prior is uniform with the added condition that they add up to one and with an upper limit of the water mass fraction of 50\% \citep{Thiabaud+2014,Marboeuf+2014}. We stress that the results of the internal structure modelling do depend to a certain extent on the chosen priors.

We ran two different versions of our model for TOI-561\,b. Given its high equilibrium temperature, any water layer would have evaporated and formed a thick water vapour atmosphere. However, with the current version of our model, it is not possible to include such a critical steam layer. We therefore choose, on the one hand, to run a dry model of the planet without an added water layer. On the other hand, we also run the full version of the model with an included water layer, as this is the most general way to model the planet that our model allows and any additional assumptions would in the end just be mirrored in the resulting posteriors. The resulting differences in the posteriors of the internal structure parameters are summarised in Table \ref{tab:int_struc_water}.

\begin{figure}
	\centering
	\includegraphics[width=0.5\textwidth]{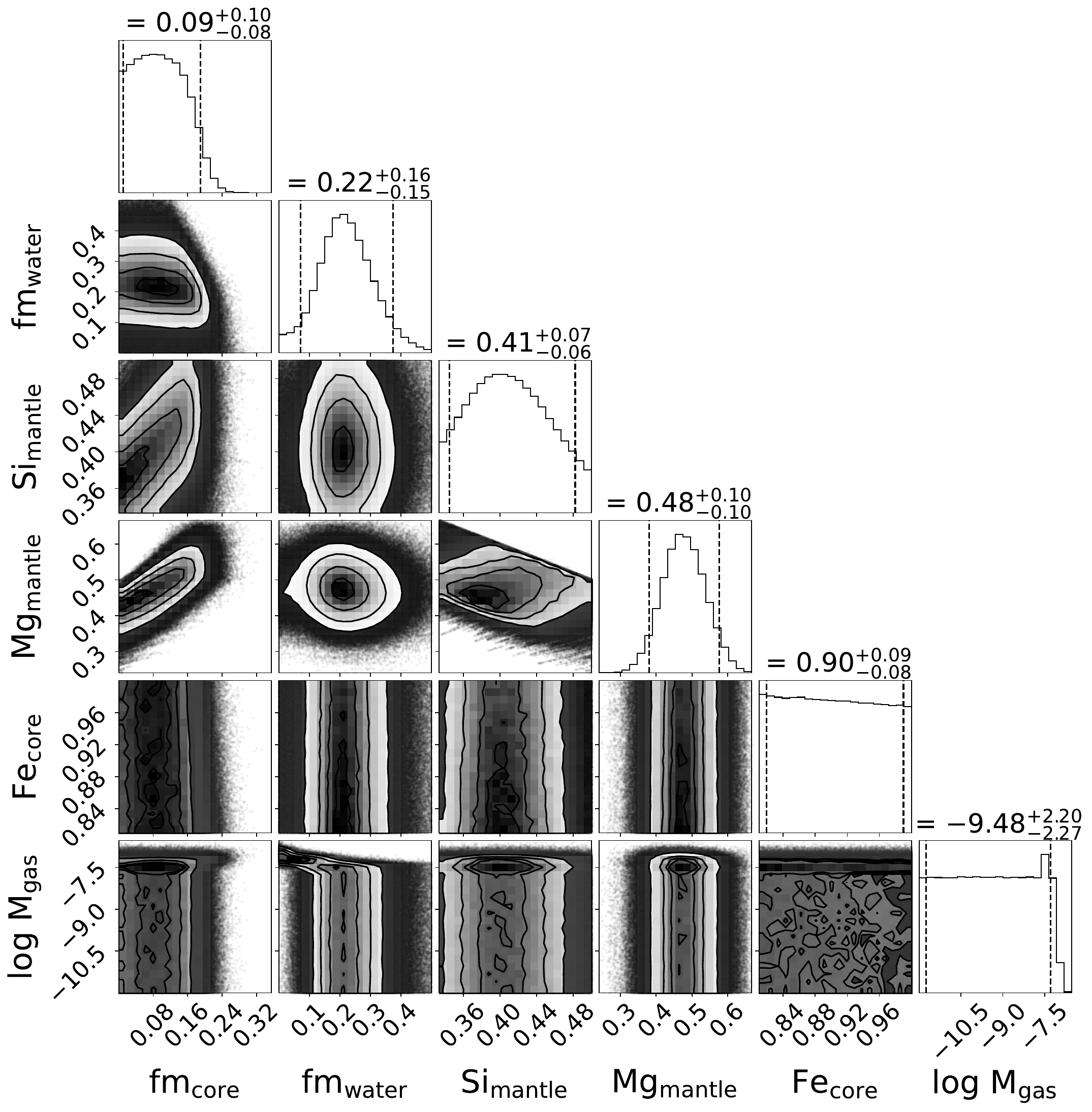}
	\caption{Corner plot of the posteriors of the internal structure parameters of TOI-561\,b. Shown are the mass fractions with respect to the solid planet of the inner core and water layer, the molar fractions of Si and Mg in the mantle layer and Fe in the inner core and the gas mass in Earth masses in a logarithmic scale. The titles on top of each column are the median and the 5 and 95 percentiles of each distribution, which are also shown by the dashed lines.}
	\label{internal structure corner plot}
\end{figure}
 
For the dry model, the posterior of the gas mass is quite well constrained with a median at around $10^{-7}$~M\textsubscript{$\oplus$}. However, such a planet is unphysical, as any H/He envelope would get evaporated very fast. Figure \ref{internal structure corner plot} shows the resulting posterior distributions of the internal structure parameters of TOI-561\,b when running the full version of the model. For both versions of the model, a planet that only consists of an iron core and a silicate mantle seems to be unlikely given the observations. The best model converges toward a water mass fraction that is constrained quite well, while the gas mass fraction is negligibly small, as expected from the strong irradiation of the planet. This finding and our core mass fraction is in agreement with a recent independent study that found a water- and gas-devoid planet can only reproduce the observed density with a composition much lower than the stellar value and the inclusion of a high melt fraction \citep{Brinkman2023}. 

We point out that in our model, the inner layers of the planet are not influenced by the pressure or temperature of the atmosphere, which does have an influence on the modelled radius of the solid planet. To investigate the influence a higher temperature could have on the radius of the iron core and silicate mantle, we selected a sub-sample of points from our posterior distribution and used the forward model to re-compute their radius when setting the temperature at the boundary between the atmosphere and the solid planet to the equilibrium temperature of TOI-561\,b. This showed that raising the temperature of the iron core and the silicate mantle only results in an increase in radius of less than 3\%, which is not enough to explain the difference in radius between the observations and a planet consisting only of an iron core and a silicate mantle.

Choosing different constraints for the Si/Mg/Fe ratios of the planet will also influence the radius of the iron core and silicate mantle. As an extreme case, we used our forward model to calculated the thickness of these two layers for an iron-free planet, again with a temperature equal to the equilibrium temperature of TOI-561\,b at the outer boundary. This gives a radius very close to the one derived for TOI-561\,b. Since a planet that is completely iron-free seems unlikely, we conclude that some heavier elements are in fact necessary. Additionally, we also performed another Bayesian analysis where we lifted the compositional constraints on the planet by allowing the code to sample from a wide range of stellar abundances, i.e. [Si/H]=[Mg/H]=[Fe/H]={\small$0^{+1}_{-1}$}. The posteriors of the internal structure parameters resulting from this analysis are summarised in Table \ref{tab:int_struc_comp}. For the Si/Fe and Mg/Fe ratios of the planet, the model gives posteriors of Si/Fe = $2.45^{+3.06}_{-1.75}$ and Mg/Fe = $2.98^{+4.18}_{-2.51}$, while the corresponding ratios derived from the stellar abundances are Si/Fe = $1.74^{+0.42}_{-0.39}$ and Mg/Fe = $2.02 \pm 0.51$. Also with this version of the model, a pure iron-silicate structure seems to be unlikely given the observations.

\begin{figure}
	\centering
	\includegraphics[width=0.5\textwidth]{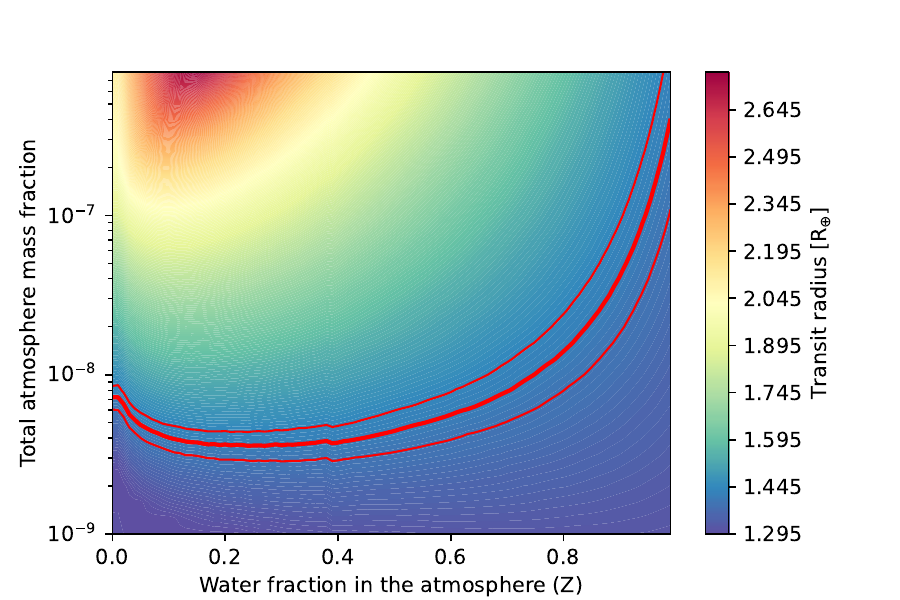}
	\caption{Transit radius as a function of the total envelope mass fraction and the molar water fraction in the envelope for a planetary structure with a fixed core mass and composition in agreement with the values for a dry model in Table \ref{tab:int_struc_water}. The red lines show the radius values that are in agreement with the measurements for TOI-561\,b.}
	\label{internal structure new analysis}
\end{figure}

However, this analysis still assumes the water layer to be in a condensed state and fully distinct from the H/He atmosphere. We therefore ran an additional exploratory internal structure analysis using a new version of our model that is fully self-consistent and features an H$_2$O enriched envelope. More specifically, we assumed the planet to have an inner iron core and a silicate mantle with a fixed combined mass of $2.24\pm0.20$ M$_{\oplus}$ and a composition corresponding to the median of the posterior distribution obtained using the dry model without water layer described above. This assumption is justified since the gas mass we obtain when running a dry model is very small (see Table \ref{tab:int_struc_water}). Note that this new model version uses the atmosphere model of \cite{Parmentier+Guillot2014} instead of the one of \cite{Lopez+Fortney2014} used in the remaining analysis. 
Then, we ran a grid of planetary forward models consisting of this fixed core and gas envelopes with different masses (going from mass fraction of $10^{-9}$ up to $10^{-6}$) and different envelope enrichments (with molar water fractions between 0 and 1). The results of this study can be seen in Fig. \ref{internal structure new analysis}. For each grid point, the transit radius of the resulting structure is shown (in color), with the red line depicting the measured radius of TOI-561\,b. Any combination of envelope mass and enrichment $Z$ that lies within the two other red lines is consistent with the measurements.

\section{Prospects of an atmosphere}\label{sec:atmos}
\subsection{Interpreting the eclipse signal}\label{subsec:pandexo}

\begin{figure*}
    \centering
    \includegraphics[width=1.5\columnwidth]{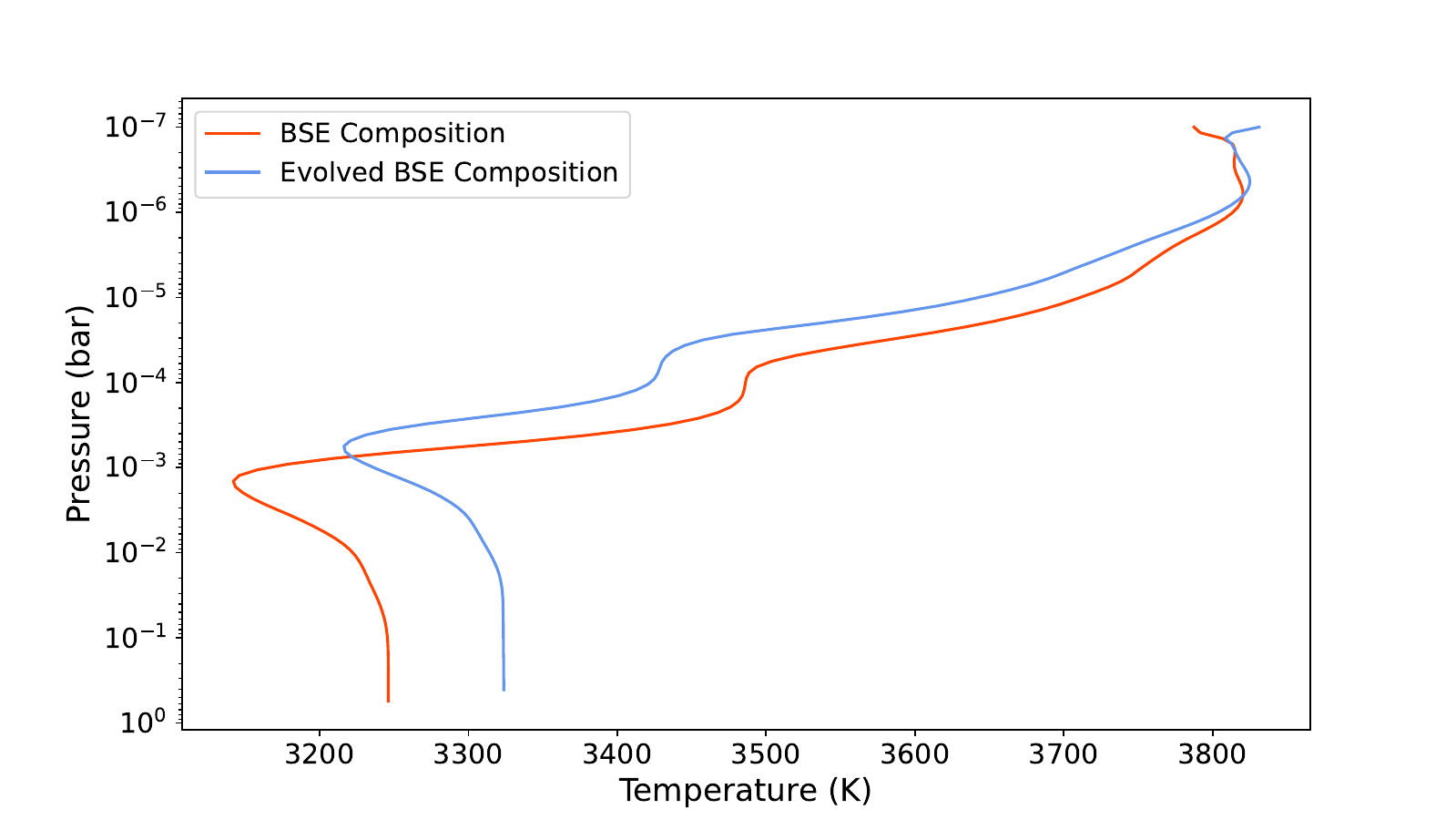}
    \includegraphics[width=1.5\columnwidth]{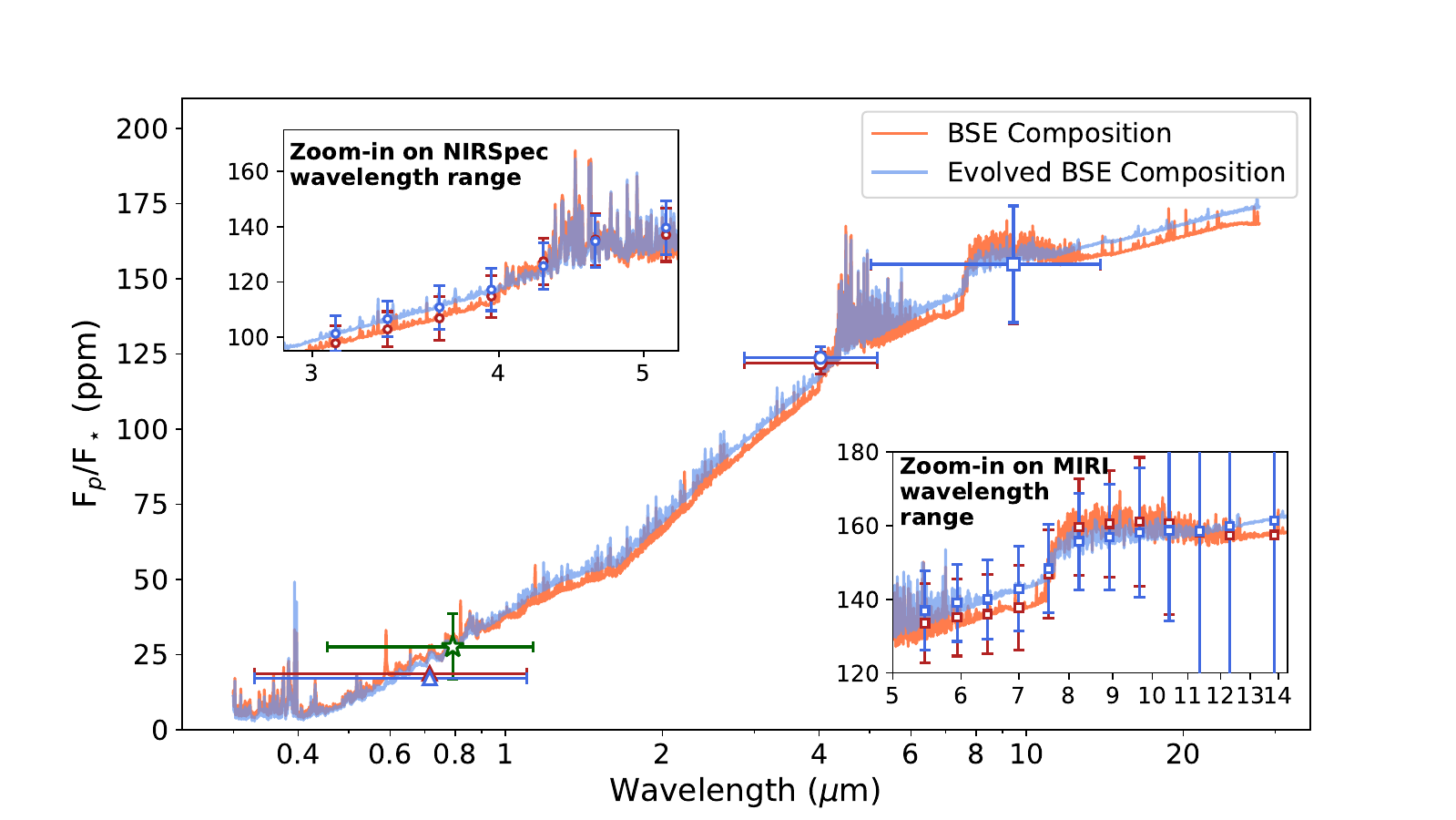}
    \caption{Theoretical models of outgassed atmosphere of TOI-561\,b. \textit{(a)} Temperature-Pressure profile of the planet for a Bulk-Silicate-Earth (BSE) composition (in orange) and an evolved BSE composition (in blue). \textit{(b)} Theoretical emission spectra for BSE (orange curve) and evolved BSE (blue curve) composition. The TESS observation of the eclipse depth is shown as a green star, where errorbars on the wavelength axis show the extent of the TESS bandpass. Other data points are simulated observations for NIRSpec (circles), MIRI (squares) and CHEOPS (triangles) for both models (represented by the colors). The white-light eclipse depths are shown for NIRSpec and MIRI with errors showing their wavelength coverage. Two inset plots display the zoom-in version of the main plot for NIRSpec and MIRI wavelength ranges. The spectroscopic simulated eclipse depths (at $R=7$) are also shown for both instruments in these plots.}
    \label{fig:pandexo}
\end{figure*}

The absence of thick ($>0.1$~bar) atmospheres on the inner two TRAPPIST-1 planets as observed by JWST \citep{Zieba2023, Greene2023} and even on a temperate rocky world like LHS 3844\,b as evidenced by Spitzer \citep{2019Natur.573...87K} appears at a first glance to make the presence of an atmosphere on a USP like TOI-561 b even less likely.

USPs, however, have such large surface temperatures ($>2000$~K) that their surfaces are expected to be at least partially molten giving rise to tenuous, evaporating atmospheres of more exotic compositions than on rocky planets in the Solar System. A tenuous rock vapour atmosphere was e.g.\ suggested on K2-141\,b based on Kepler and Spitzer observations \citep{2022arXiv220300370Z}. Thus, constraining such evaporating atmospheres on USPs raised the tantalizing prospect to assess surface properties on rocky exoplanets more directly by observing spectral features of outgassed molecules \citep[see e.g.][]{Chao2021}

Recently, \citet{2022A&A...661A.126Z} comprised an extensive catalogue of outgassed atmospheres for various USPs, including TOI-561\,b, under different assumptions for surface composition and outgassing efficiency. These authors provided pressure temperature profiles for these different atmospheres that were calculated self-consistently, assuming irradiation at the substellar point without any heat distribution and assuming an albedo of 0. In the case of TOI-561\,b, the model yields about 0.4 bar for 60\% outgassing efficiency.

We use two planetary compositions, bulk silicate (oxidized) Earth (BSE) and evolved BSE (with 60\% outgassing efficiency), to estimate the eclipse depth in the TESS bandpass to explain the measured eclipse signal described in Section \ref{subsec:eclipse}\footnote{Models are publicly available at \url{https://github.com/zmantas/LavaPlanets}}. These two models yield SiO as the main volatile in the outgassed atmosphere. The predicted eclipse depths for these models are 27.81\,ppm and 26.59\,ppm, which is very close to the observed eclipse depth of 27.40\,ppm. This points to the fact that the emission coming from the planet in the TESS bandpass is the result of thermal emission only. However, we note here that models assume zero albedo and no heat re-distribution resulting in a hotter temperature profile (see, e.g., Figure \ref{fig:pandexo} (a)) which was already noted by \citet{2022A&A...661A.126Z} as a limitation of their methods. It has been pointed out, however, that lava surfaces should have very low albedos \citep{Essack2020}. Further, K2-141\,b, the only USP so far with hints of an outgassed rock vapour atmospheres, exhibits a phase-curve consistent with a scenario without any heat re-distribution.

Keeping in mind these caveats, we make predictions for potential CHEOPS and upcoming JWST observations of this planet. The BSE and evolved BSE models predict the eclipse depth of 18.74\,ppm and 17.18\,ppm in the CHEOPS bandpass (orange and blue triangles in Figure \ref{fig:pandexo} (b)). A larger observed depth in the CHEOPS bandpass could be a sign of a significant reflective component. However, it would take an impractically large number of observations (62 and 73 for two models) to constrain the above-mentioned eclipse depths at 3$\sigma$ as the target is faint with at $G = 10.0$\,mag\footnote{We used CHEOPS Exposure Time Calculator (\url{https://cheops.unige.ch/pht2/exposure-time-calculator/}) to estimate this.}. Furthermore, it is questionable if lava worlds exhibit large albedos, at least if an oxidized silicate surface like in the BSE scenario is assumed \citep{Essack2020}. It would take ultramafic surfaces composed of olivine and enstatite to assume larger albedos (>0.2) \citep{Hu2012}.

On the other hand, the James Webb Space Telescope (JWST) would be ideal to observe the thermal emission from the planetary surface or its atmosphere. Indeed, a Cycle 2 GO program (GO 3860, PI: J. Teske) aims to do the same by observing phase curves of the planet with NIRSpec/G395H. Here we use \citet{2022A&A...661A.126Z} models with PandExo \citep{2017PASP..129f4501B} to simulate observations for NIRSPec/G395H and MIRI/LRS. NIRSpec/G395H instrument, which has a wavelength range of $2.87-5.18\,\mu$m, covers a SiO feature between $4\,\mu$m and $5 \ \mu$m. With 4 occultations of the planet (as asked for in GO 3860) we can expect to detect this feature as a rise in eclipse depth after $\sim 4\,\mu$m as shown in upper-left inset panel of Figure \ref{fig:pandexo} (b) (orange and blue circles). Unfortunately, both BSE and evolved BSE models are very similar in this range. Therefore the NIRSpec/G395H observations cannot distinguish between both models. There is an additional SiO feature near $\sim 9\,\mu$m covered by MIRI/LRS wavelength range, $5.02 - 13.86\,\mu$m. This feature is interesting because the amplitude (along with the baseline) of the feature changes depending on the outgassing efficiency. As shown in Figure \ref{fig:pandexo} (b), the evolved BSE composition (in blue) has a lower amplitude and high baseline of the feature compared to that of the BSE composition (in orange). Additionally, there is a small SiO$_2$ feature near $\sim 7\,\mu$m which is only present in spectrum of BSE composition (see the lower-right inset panel in Figure \ref{fig:pandexo} (b)). Both of these features can be helpful in distinguishing the two models. We, however, note that with the current precision of the MIRI/LRS instrument, it would be quite challenging to reach the required level of noise floor to differentiate two models at higher statistical confidence. We used 10 eclipse observations for our simulation here.

The white-light eclipse depth could be used to estimate the temperature of the planet and solve the degeneracy between thermal and reflective components (also shown in Figure \ref{fig:ag_vs_temp}). As already mentioned, models from \citet{2022A&A...661A.126Z} suggest that the emission in the TESS bandpass is only because of thermal radiation. If this is the case, then the white-light eclipse depth should be around 121.9 and 123.7 ppm for NIRSpec/G395H and 154.7 and 154.8 ppm for MIRI/LRS, respectively for BSE and evolved BSE compositions (see, Figure \ref{fig:pandexo} (b)). A lower observed white-light eclipse depths in these bands would be a hint of lower dayside temperature and a non-zero bond albedo.

\subsection{Search for variability}\label{subsec:variability}

\begin{figure}
    \centering
    \includegraphics[width=\columnwidth]{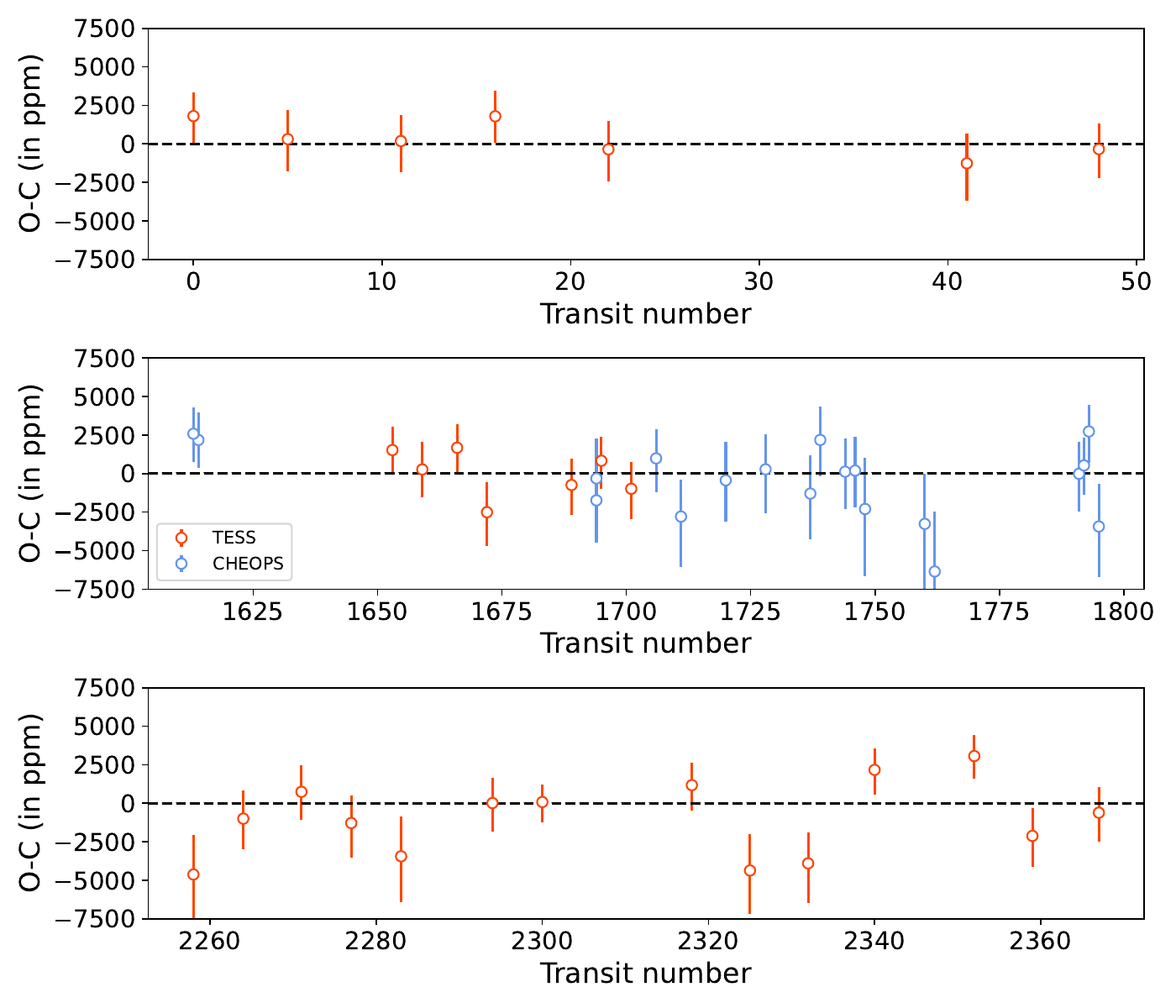}
    \caption{Difference in ppm between the $R_p/R_\star$ measured in individual (for CHEOPS data) or group of five (for TESS data) transits and the value of $R_p/R_\star$ from the joint CHEOPS-TESS analysis. The orange and blue points are the results from TESS and CHEOPS analysis, respectively.}
    \label{fig:oc_dep}
\end{figure}

As mentioned in Section \ref{subsec:eclipse} the dayside temperature of the planet can reach as high as $\sim 3000$\,K, resulting in the surface of the planet being partially or completely molten. At this point, the planet can support a secondary outgassed atmosphere made up of heavier species such as Na, O$_2$, SiO etc. \citep{2009ApJ...703L.113S, 2011ApJ...742L..19M, 2012ApJ...755...41S, 2021arXiv211204663W}. Driven by winds, the species could travel to the nightside and condensate, or even escape into space \citep{2016ApJ...828...80K, 2021arXiv211204663W}. Both phenomena, influenced by the surface activity of the planet and the stellar irradiation, could lead to temporal variability in the atmospheric properties, which could potentially be imprinted on the $R_p/R_\star$ measured over several transit events.

Motivated by this rationale, we re-analysed all of the TESS and CHEOPS data, but now fixed all parameters except $R_p/R_\star$ to the values derived from our joint photometric analysis (Table \ref{tab:fitted_transit_param}). While for the CHEOPS data we model each transit individually, for TESS we group $\sim 5$ transit events together to achieve a better signal-to-noise ratio. The result of this analysis is presented in Figure \ref{fig:oc_dep}. It shows the difference in ppm between each measured $R_p/R_\star$ and the $R_p/R_\star$ from our joint analysis. As can be seen from Figure \ref{fig:oc_dep}, we find no significant variations of $R_p/R_\star$ over time. Although this does not rule out the possibility of the atmosphere being variable, our current dataset does not have the required precision.

\section{Discussion and conclusions}\label{sec:conclusions}

\begin{figure*}
    \centering
    \includegraphics[width=1.5\columnwidth]{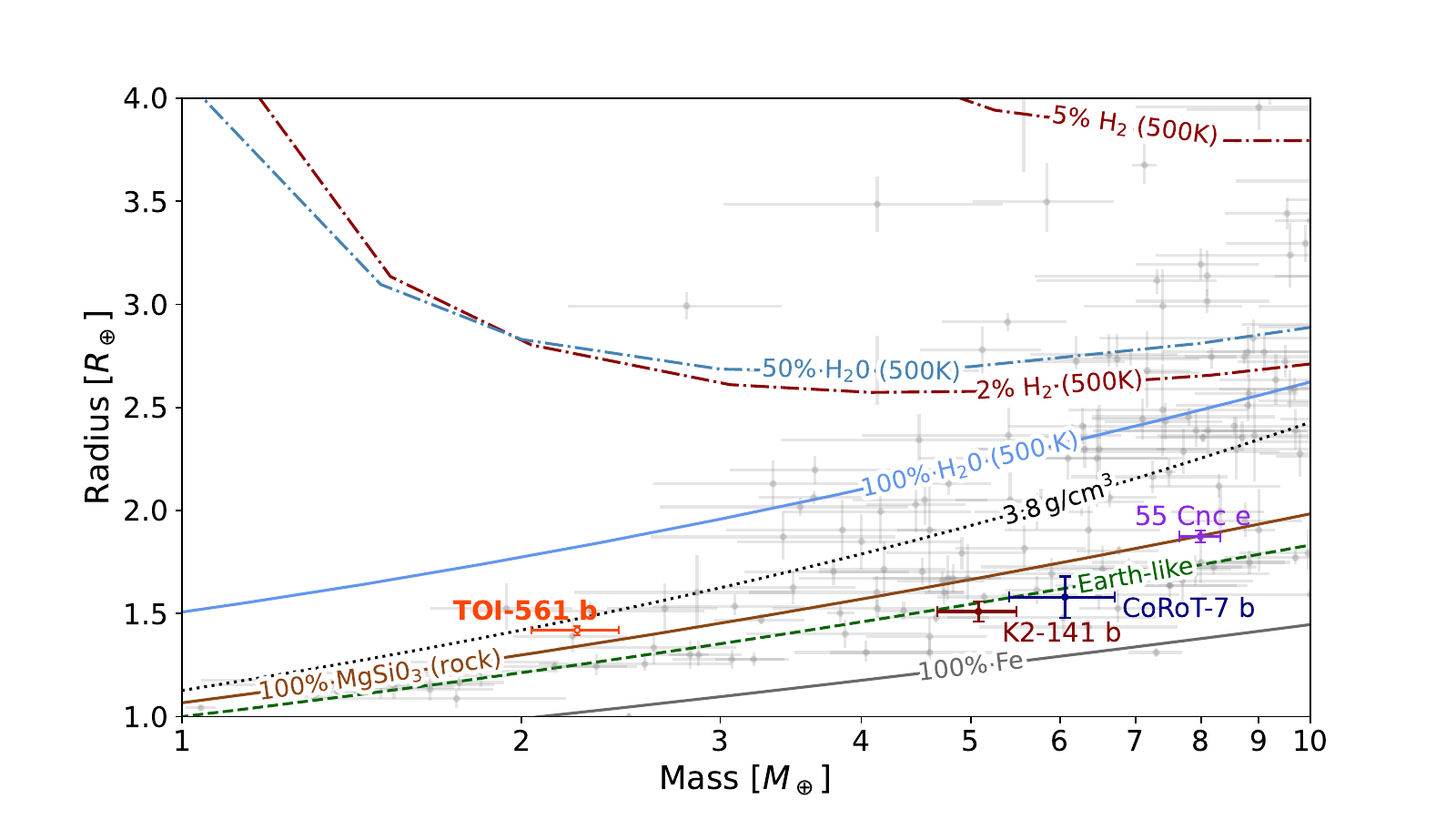}
    \caption{Mass-radius diagram of exoplanets with $M_p \leq 10\,M_\oplus$ and $R_p \leq 4\,R_\oplus$ known with a precision better than 30\%. Gray points are all known exoplanets as of June 19, 2023 taken from the \citet{PSCompPars}. The black dashed line is the iso-density curve for the value we derived for TOI-561\,b. Mass of TOI-561\,b is taken from \citet{Brinkman2023} and radius from the present work. Also shown are theoretical models for various chemical compositions as computed by \citet{zeng_2019}. Mass and radius of 55 Cnc\,e, K2-141\,b and CoRoT-7\,b are taken from \citet{2018A&A...619A...1B}, \citet{2018AJ....155..107M} and \citet{2022MNRAS.515.3975J}.}
    \label{fig:mr}
\end{figure*}

USPs are a singular class of planets orbiting their host star within a day. This property makes them appealing targets to study various compelling phenomena caused by the extreme proximity and radiation from the star, ranging from their internal structure to their atmospheres. Indeed, well-known USPs like 55 Cnc\,e, K2-141\,b, or LHS 3844\,b display a diverse range of composition from a thick, possibly variable atmosphere to no atmosphere at all \citep{2016MNRAS.455.2018D, 2019Natur.573...87K, 2022arXiv220300370Z}. Among the USP population, the newly discovered TOI-561\,b stands apart by its unusually low density of {\small $4.3049 ^{+ 0.4411} _{-0.4216}$\,g/cm$^3$} when other USPs show comparatively higher densities (see the position of TOI-561\,b in the Mass-Radius diagram in Figure \ref{fig:mr}).

\begin{figure}
    \centering

    \includegraphics[width=\columnwidth]{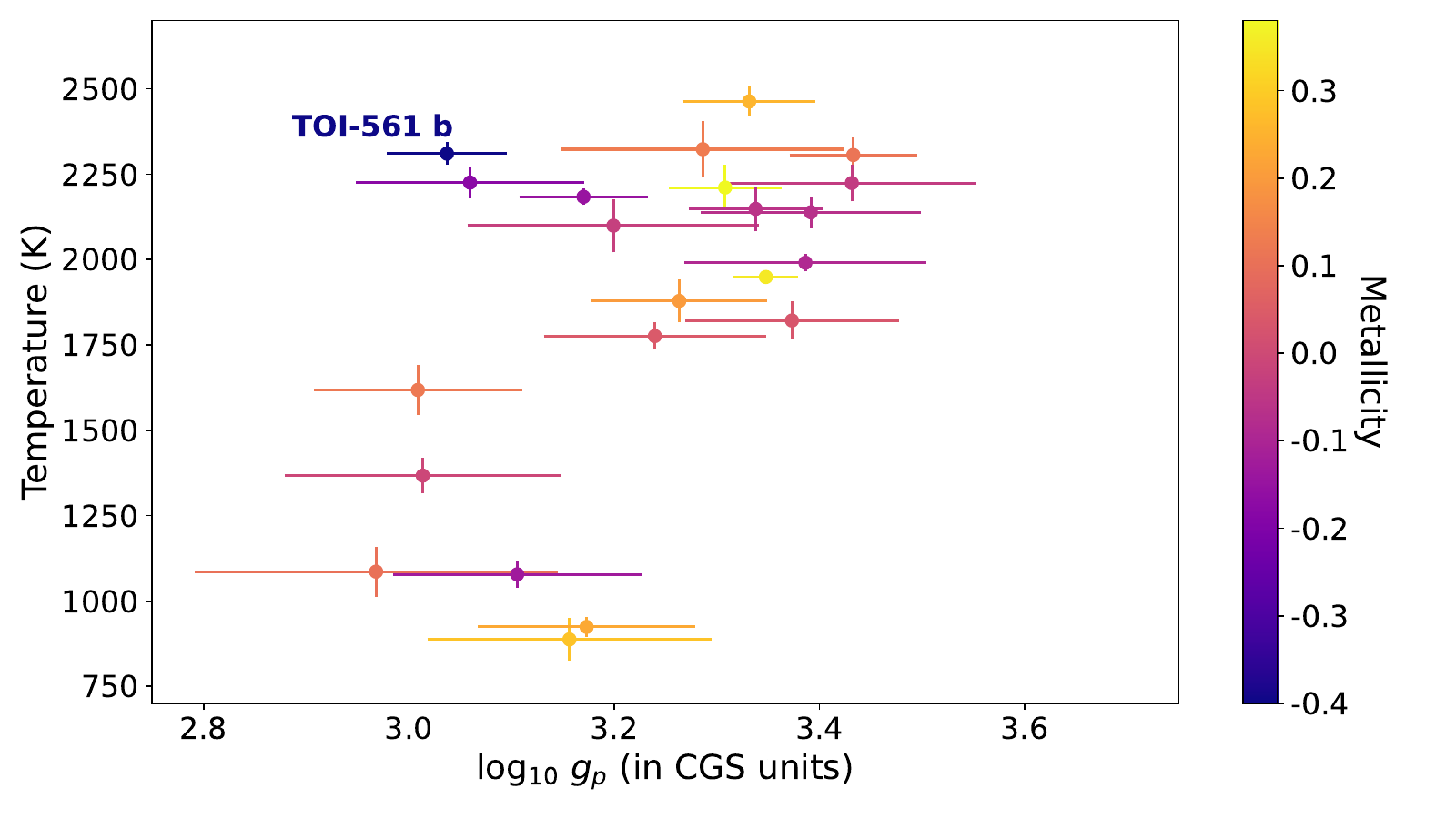}
    \includegraphics[width=\columnwidth]{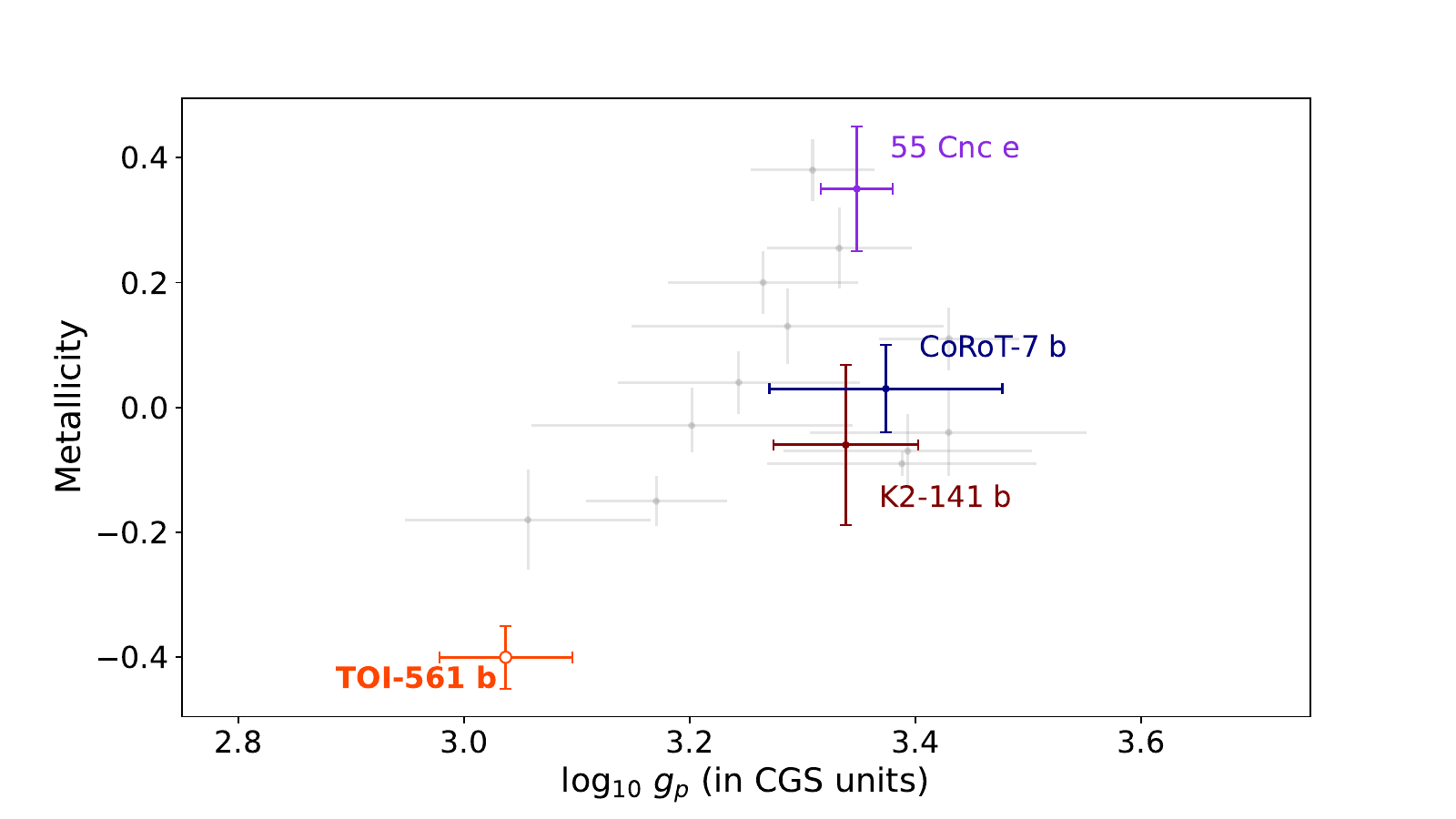}
    \caption{Population of USPs in (\textit{a}) planetary surface gravity - equilibrium temperature and (\textit{b}) host star metallicity - surface gravity spaces. In the top panel, the data points are color-coded with the metallicity of the host star. The USP population is defined as all confirmed planets with an orbital period of less than a day and $M_p \leq 10 \ M_\oplus$. The data of the systems is taken from \citet{PSCompPars} on June 19, 2023. Additionally, the bottom planets only include planets hotter than 1700 K (\textit{see, text}).}
    \label{fig:demographics}
\end{figure}

To further place TOI-561\,b in the context of known USPs and understand how the metal-poor nature of the host star may have affected its formation and evolution we retrieve planet properties for all well-characterised bodies with an orbital period of less than a day and $M_p \leq 10 \ M_\oplus$ from \citet{PSCompPars}. To test how stellar composition may impact USPs, we first make the assumption that these bodies have entirely lost their primordial Hydrogen atmosphere and as such, we are only observing the refractory components of the planets. To assess this assumption, we computed the minimum mean molecular weight (MMMW) of a potential atmosphere of our USP sample using a formulation of the Jeans escape criteria \citep{Jeans1925} that includes Roche lobe effects \citep{Erkaev2007,Fossati2017} in a similar manner to \citet{2022MNRAS.511.1043W}. Given their current physical and orbital parameters, we find that all USPs with the exception of TOI-1075\,b \citep{Essack2023} cannot have held onto a primordial envelope and thus our assumption is valid. We recognise that other processes may have induced further atmospheric mass loss, and whilst the inclusion of this additional modelling is beyond the scope of the paper, any additional atmospheric escape would strengthen our assumption that the USPs are primarily rocky. Furthermore, we find that the minimum mean molecular weight of an atmosphere that TOI-561\,b can retain is 11\,amu, providing evidence that the planet has likely lost its primordial atmosphere, but may have a heavier, secondary envelope.

Assured that the planets in our USP sample do not have primordial atmospheres, we assess the physical properties of these well-characterised bodies to probe underlying processes. In the upper panel of Figure \ref{fig:demographics}, we plot the planetary surface gravity of rocky USPs as determined by our MMMW analysis above against their equilibrium temperature that reveals a clear trend. To substantiate the identified correlation, we used a Bayesian correlation tool \citep{Figueira2016} and find the median and standard deviation of the correlation metric distribution (similar to Spearman’s rank value) to be 0.45$\pm$0.16. This strong trend is interpreted as lower surface gravity planets have lower gravitational potential wells that allow for the escape of envelopes at lower temperatures and hence appearing as the rocky USPs identified by our MMMW analysis. \citet{Adibekyan+2021} found evidence for stellar composition influencing planetary internal structure for terrestrial bodies. If this finding is universal, one would expect planets around metal-poor stars to have smaller iron cores and thus lower surface gravities for bodies of a similar radius. Interestingly, if we remove metal-poor planets, [Fe/H] $<$ -0.1, (TOI-561\,b, Kepler-10\,b, Kepler-78\,b, and TOI-1685\,b) from our USP sample, the correlation between surface gravity and equilibrium temperature increases to 0.65$\pm$0.13. This is because metal-poor USP planets have lower surface gravities at similar equilibrium temperatures. This supports previous findings that host star metallicity alters planet interior structure \citep{Adibekyan+2021,2022MNRAS.511.1043W}. This is highlighted in the lower panel of Figure \ref{fig:demographics} by removing the equilibrium temperature dependence by selecting USPs hotter than 1700\,K and finding a strong trend between surface gravity and host-star metallicity.

Figure \ref{fig:demographics} also clearly shows the uniqueness of TOI-561\,b in temperature-metallicity-surface gravity space. All these motivated our follow-up of this planet with the state-of-the-art photometric telescope CHEOPS to measure precisely its radius, and therefore constrain its composition and structure.
We acquired 13 new transit observations of the planet with CHEOPS, and combined them with three archival visits from CHEOPS and four TESS sectors. This joint dataset allowed us to set strong constraints on the planet-to-star radius ratio, reducing its uncertainty to $\sim 2\%$. We used our updated parameters for TOI-561\,b to model its internal structure. We find that the structure of planet is consistent with negligible primary volatile atmosphere made up of H/He which is expected from the atmospheric escape from such a highly irradiated planet. Additionally, a planet built only from an iron core and silicate mantle is not enough to explain the observed density, supporting the presence for lighter material in the planetary structure. This could be a thin layer of high mean molecular weight (e.g. a water layer, as shown by our internal structure modelling), or a larger silicate mantle than what is predicted using a one-to-one scaling between the stellar and planet composition, or a combination of both. We also ran a fully self-consistent exploratory forward model with water enriched envelope and demonstrate that such a model can indeed explain the transit radius. We show that a range of water fraction in the atmosphere is possible for a varying atmospheric mass fraction. Given the high equilibrium temperature of the planet, a plausible scenario is that it sustains a secondary atmosphere containing heavy metal species. We searched for variations in $R_p/R_\star$ over time, which could trace the variability expected from such an envelope, but found no such evidence.

In addition to the transit observations, we also find a weak detection of the secondary eclipse in the TESS data, with an eclipse depth {\small $L=27.40 ^{+10.87} _{-11.33}$\,ppm}. Since this emission signal would be contaminated from the reflective and thermal radiation from the planet we cannot uniquely determine its geometric albedo and dayside temperature. The measured occultation depth would correspond to a dayside temperature of $\sim 3325$\,K if the signal is assumed to originate entirely from thermal radiation; on the contrary, a hundred per cent contribution from reflection would imply a geometric albedo of $\sim 0.83$. Given that the planet could possibly have a secondary metal-rich atmosphere, we use models of outgassed atmospheres from \citet{2022A&A...661A.126Z} to explain this eclipse signal. The expected eclipse depths from the two model compositions, bulk silicate (oxidized) Earth (BSE) and evolved BSE composition with 60\% outgassing efficiency, are 27.81 ppm and 26.59 ppm, which is very close to the observed value. This implies that the emission in TESS bandpass is mainly of thermal origin.

The composition of the atmosphere is essential in constraining the surface properties, given the possible atmosphere-interior interactions. Our photometric observations cannot constrain the presence of a specific species in the atmosphere of the planet. While ground-based spectroscopy could be used to search for light species escaping the planet, such as sodium or oxygen, space-based observations in the infrared could allow us to detect a mineral atmosphere. The recently launched James Webb Space Telescope (JWST) would be the ideal facility to search for dust and metals in the atmosphere of TOI-561\,b, which is one of the main goals of the recently accepted Cycle 2 GO program (GO 3860, PI: J. Teske). Using the models from \citet{2022A&A...661A.126Z} we anticipate that these observations, which will use NIRSpec/G395H instrument, should be able to detect SiO in the atmosphere. Observations in mid-infrared using MIRI/LRS (mode not used by GO 3860) could, in principle, not only detect the mineral species in the atmosphere but also infer the surface evolution. Infrared observations have the other advantage of breaking the degeneracy between reflective and thermal contributions in the observed emission. Hence, with the help of observations at longer wavelengths, we can not only identify the composition of the atmosphere, but also derive the temperature of the planet.

\begin{acknowledgements}


We would like to thank an anonymous referee for their detailed referee report and suggestions which significantly improved the manuscript.

CHEOPS is an ESA mission in partnership with Switzerland with important contributions to the payload and the ground segment from Austria, Belgium, France, Germany, Hungary, Italy, Portugal, Spain, Sweden, and the United Kingdom. The CHEOPS Consortium would like to gratefully acknowledge the support received by all the agencies, offices, universities, and industries involved. Their flexibility and willingness to explore new approaches were essential to the success of this mission. 

This research has made use of the NASA Exoplanet Archive, which is operated by the California Institute of Technology, under contract with the National Aeronautics and Space Administration under the Exoplanet Exploration Program.

JAP would like to thank Yamila Miguel for an insightful discussion on the atmosphere of lava planets and for kindly providing theoretical models for our planet.

JAP and ABr were supported by the SNSA.

JAE and YA acknowledge the support of the Swiss National Fund under grant 200020\_172746.

This project has received funding from the European Research Council (ERC) under the European Union's Horizon 2020 research and innovation programme (project {\sc Spice Dune}, grant agreement No 947634).

ACC and TGW acknowledge support from STFC consolidated grant numbers ST/R000824/1 and ST/V000861/1, and UKSA grant number ST/R003203/1.

LCa acknowledges financial support from the Österreichische Akademie der Wissenschaften; LCa acknowledges support from the European Union H2020-MSCA-ITN-2019 under Grant Agreement no. 860470 (CHAMELEON).

S.G.S. acknowledge support from FCT through FCT contract nr. CEECIND/00826/2018 and POPH/FSE (EC). 
This project has received funding from the European Research Council (ERC) under the European Union's Horizon 2020 research and innovation programme (project {\sc Four Aces}; grant agreement No 724427).  It has also been carried out in the frame of the National Centre for Competence in Research PlanetS supported by the Swiss National Science Foundation (SNSF). D.Eh. and A.De. acknowledge financial support from the SNSF for project 200021\_200726. 
KWFL acknowledge support by DFG grants RA714/14-1 within the DFG Schwerpunkt SPP 1992, ``Exploring the Diversity of Extrasolar Planets''.
ML acknowledges support of the Swiss National Science Foundation under grant number PCEFP2\_194576. 
We acknowledge support from the Spanish Ministry of Science and Innovation and the European Regional Development Fund through grants ESP2016-80435-C2-1-R, ESP2016-80435-C2-2-R, PGC2018-098153-B-C33, PGC2018-098153-B-C31, ESP2017-87676-C5-1-R, MDM-2017-0737 Unidad de Excelencia Maria de Maeztu-Centro de Astrobiología (INTA-CSIC), as well as the support of the Generalitat de Catalunya/CERCA programme. The MOC activities have been supported by the ESA contract No. 4000124370. 
S.C.C.B. acknowledges support from FCT through FCT contracts nr. IF/01312/2014/CP1215/CT0004. 
XB, SC, DG, MF and JL acknowledge their role as ESA-appointed CHEOPS science team members. 
This project was supported by the CNES. 
The Belgian participation to CHEOPS has been supported by the Belgian Federal Science Policy Office (BELSPO) in the framework of the PRODEX Program, and by the University of Liège through an ARC grant for Concerted Research Actions financed by the Wallonia-Brussels Federation. 
L.D. is an F.R.S.-FNRS Postdoctoral Researcher. 
This work was supported by FCT - Fundação para a Ciência e a Tecnologia through national funds and by FEDER through COMPETE2020 - Programa Operacional Competitividade e Internacionalizacão by these grants: UID/FIS/04434/2019, UIDB/04434/2020, UIDP/04434/2020, PTDC/FIS-AST/32113/2017 \& POCI-01-0145-FEDER- 032113, PTDC/FIS-AST/28953/2017 \& POCI-01-0145-FEDER-028953, PTDC/FIS-AST/28987/2017 \& POCI-01-0145-FEDER-028987, O.D.S.D. is supported in the form of work contract (DL 57/2016/CP1364/CT0004) funded by national funds through FCT. 
B.-O.D. acknowledges support from the Swiss National Science Foundation (PP00P2-190080). 
MF gratefully acknowledge the support of the Swedish National Space Agency (DNR 65/19, 174/18). 
DG gratefully acknowledges financial support from the CRT foundation under Grant No. 2018.2323 ``Gaseousor rocky? Unveiling the nature of small worlds''. 
M.G. is an F.R.S.-FNRS Senior Research Associate. 
SH gratefully acknowledges CNES funding through the grant 837319. 
KGI is the ESA CHEOPS Project Scientist and is responsible for the ESA CHEOPS Guest Observers Programme. She does not participate in, or contribute to, the definition of the Guaranteed Time Programme of the CHEOPS mission through which observations described in this paper have been taken, nor to any aspect of target selection for the programme. 
This work was granted access to the HPC resources of MesoPSL financed by the Region Ile de France and the project Equip@Meso (reference ANR-10-EQPX-29-01) of the programme Investissements d'Avenir supervised by the Agence Nationale pour la Recherche. 
PM acknowledges support from STFC research grant number ST/M001040/1. 
VNa, IPa, GPi, RRa and GSc acknowledge support from CHEOPS ASI-INAF agreement n. 2019-29-HH.0. 
This work was also partially supported by a grant from the Simons Foundation (PI Queloz, grant number 327127). 
IRI acknowledges support from the Spanish Ministry of Science and Innovation and the European Regional Development Fund through grant PGC2018-098153-B- C33, as well as the support of the Generalitat de Catalunya/CERCA programme. 
GyMSz acknowledges the support of the Hungarian National Research, Development and Innovation Office (NKFIH) grant K-125015, a a PRODEX Experiment Agreement No. 4000137122, the Lend\"ulet LP2018-7/2021 grant of the Hungarian Academy of Science and the support of the city of Szombathely. 
V.V.G. is an F.R.S-FNRS Research Associate. 
NAW acknowledges UKSA grant ST/R004838/1.
      
\end{acknowledgements}

%
%

\bibliographystyle{aa} 
\bibliography{references} 

\begin{appendix}
\onecolumn

\section{Photometric analysis without assuming prior knowledge on the stellar density}\label{app:a}
\begin{figure}[ht]
    \centering
    \includegraphics[width=11cm]{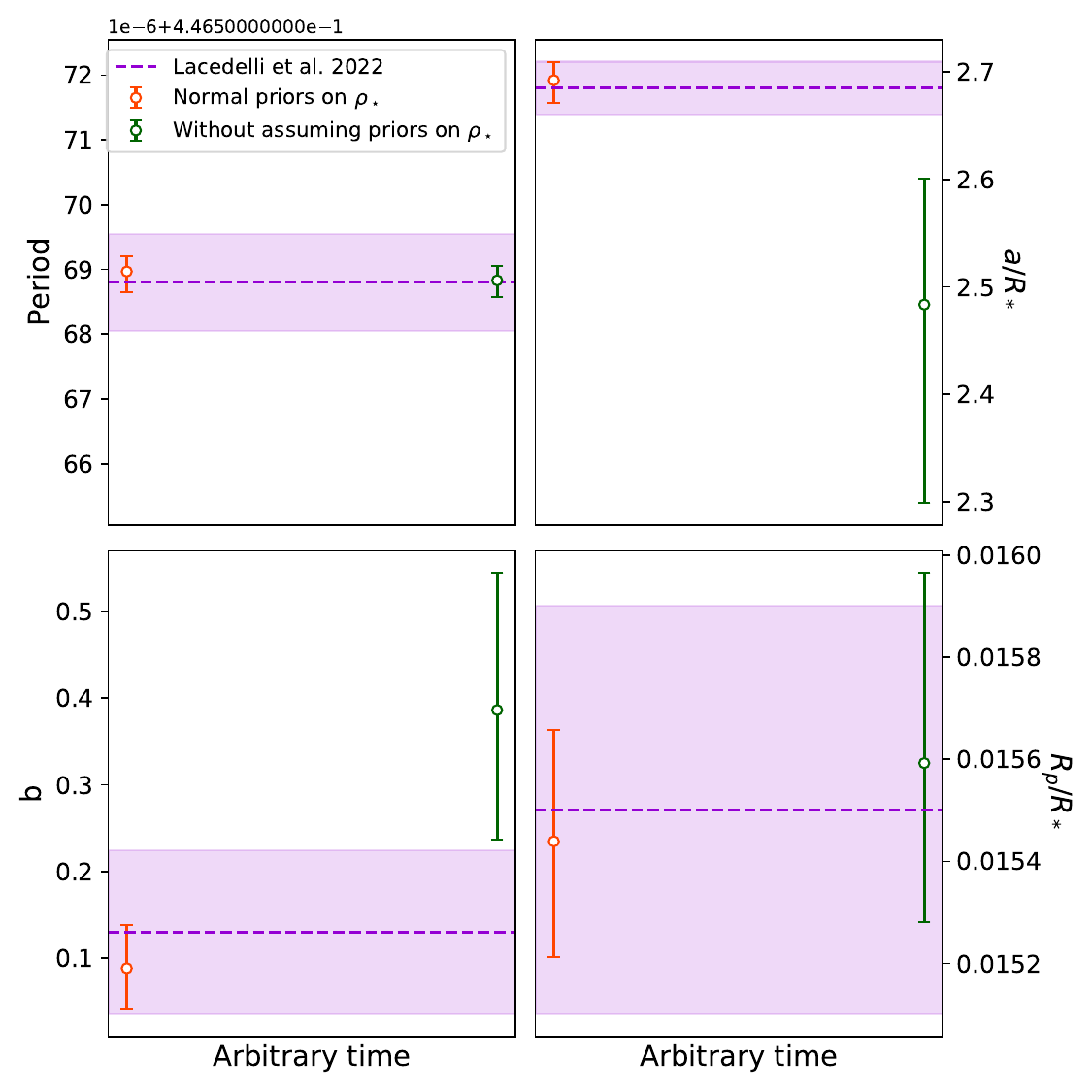}
    \caption{Comparison of some of the retrieved planetary parameters (Period $P$, scaled semi-major axis $a/R_\star$, impact parameter $b$ and the planet-to-star radius ratio $R_p/R_\star$) between our two analysis: the main analysis in which we used informative priors on the stellar density (orange points) and the analysis in which we assumed no prior knowledge on the stellar density (green points). The purple dashed line with the band show the literature values of the parameters along with uncertainties from \cite{2022MNRAS.511.4551L}. Note that the scale on the y-axis is the relative.}
    \label{fig:wo_rho}
\end{figure}

As mentioned in Section \ref{subsec:transit_ana}, in our joint transit fit for the full dataset of CHEOPS and TESS, we used informative priors on the stellar density. The stellar density adopted in the procedure was computed from the stellar mass and radius estimated from the stellar spectroscopic analysis performed in \citet{2022MNRAS.511.4551L}. For completeness, and to test the validity of our results, we perform another analysis. This time, we do not assume any prior knowledge on the stellar density. Therefore, in this analysis, we put a wide uninformative priors on $a/R_\star$ (scaled semi-major axis, uniformally distributed priors in the log-space between 1 and 10) instead of using the stellar density. We illustrate the result of this analysis in Figure \ref{fig:wo_rho} by comparing them with the analysis performed in Section \ref{subsubsec:joint_ana}. Given the precision and quantity of the data at hand, the planetary parameters agree very well (with $3-\sigma$, even $1-\sigma$ for some of them) with our analysis when we use priors on the stellar density and with their literature counterparts from \citet{2022MNRAS.511.4551L}. However, as expected, the uncertainty on the parameters are increased in this analysis owing to our lack of knowledge of semi-major axis/stellar density.

\clearpage

\section{Raw and detrended photometry from CHEOPS}\label{app:b}

\begin{figure}[ht]
    \centering
    \includegraphics[width=\textwidth]{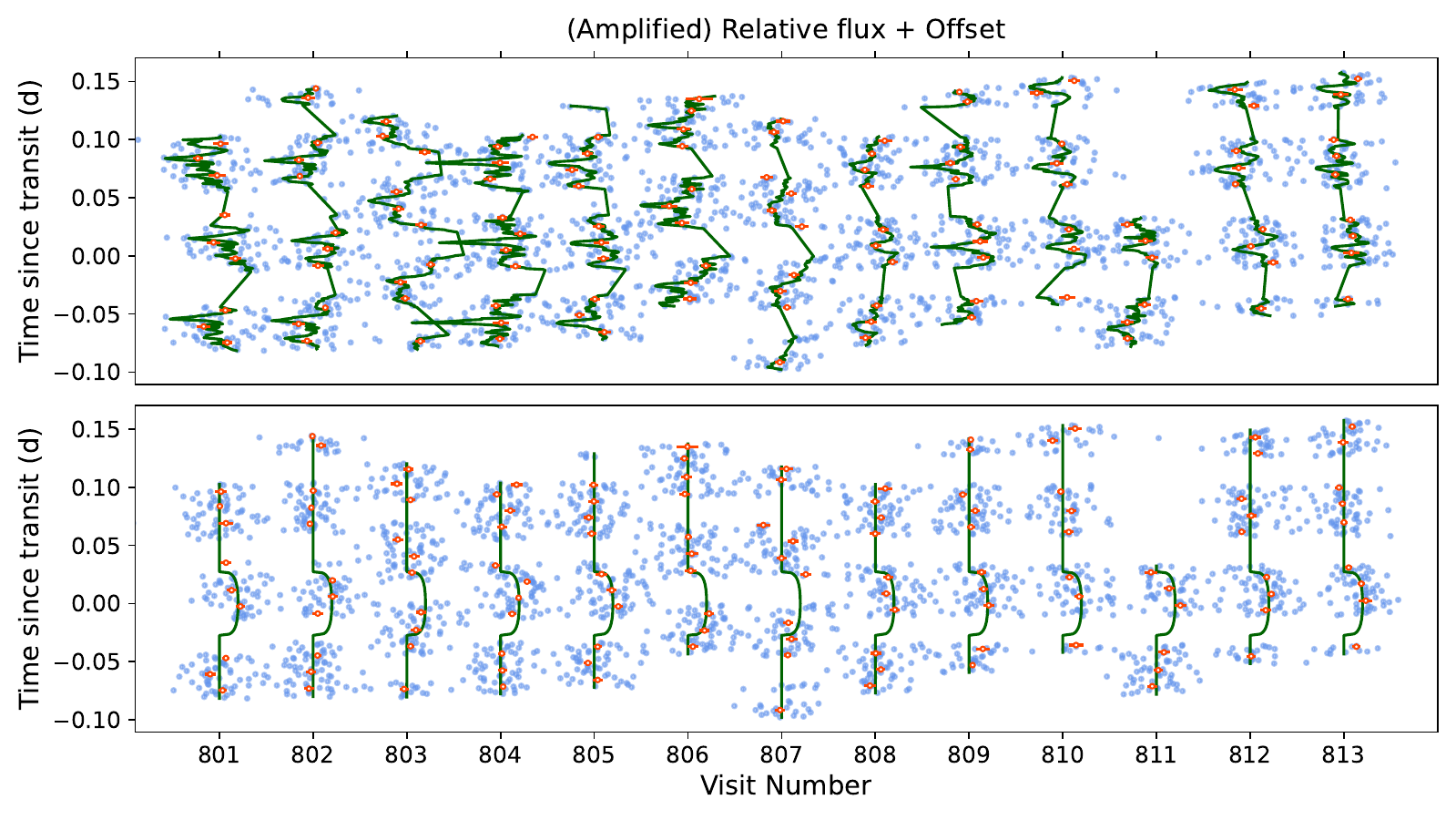}
    \includegraphics[width=\textwidth]{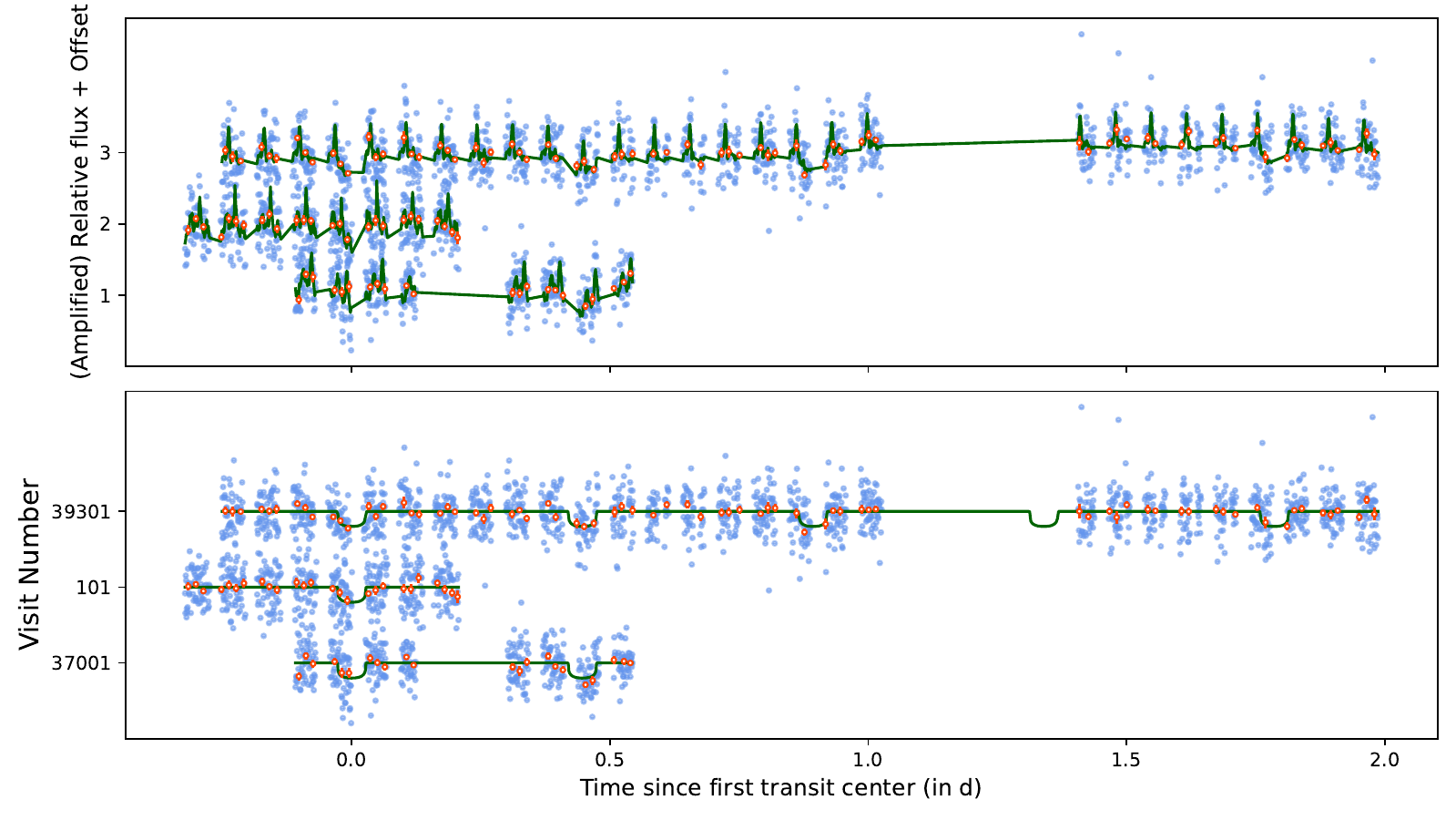}
    \caption{Raw and detrended CHEOPS observations. \textit{(a)}: Raw (top panel) and detrended photometry (bottom panel) for our new CHEOPS observations. The blue points and orange points are the original and binned data points. The fitted median full model (in top panel) and transit model (in bottom panel) is shown with dark green lines. \textit{(b)}: Same as \textit{(a)}, but for archival CHEOPS visits. Note that, for clarity, we masked the transit signals from other planets.}
    \label{fig:all_photometry}
\end{figure}

\clearpage

\section{Correlation plot of the fitted transit parameters}\label{app:c}

\begin{figure}[ht]
    \centering
    \includegraphics[width=\textwidth]{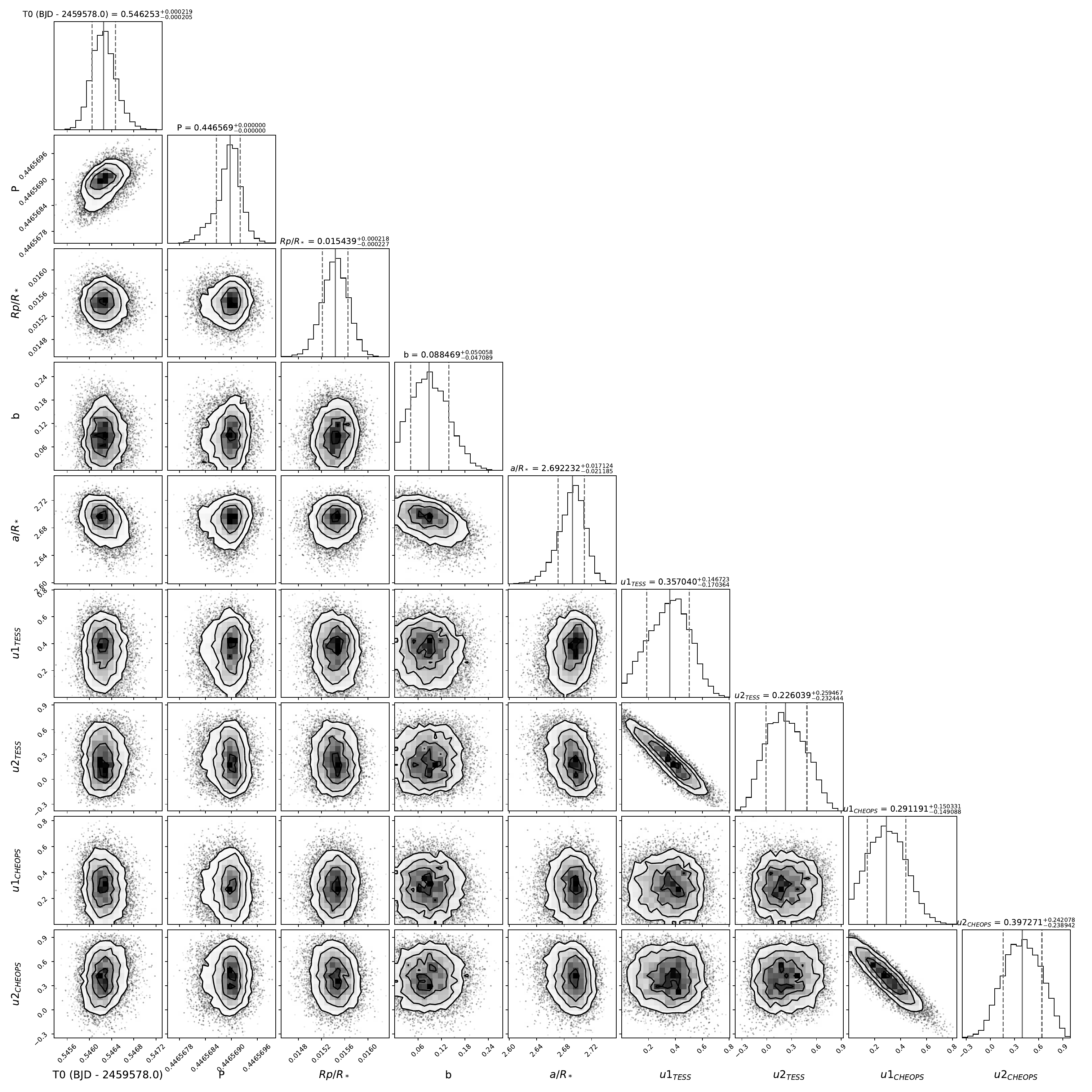}
    \caption{Correlation plot of the fitted transit parameters of planet b in our joint photometric analysis.}
    \label{fig:corner_plot}
\end{figure}

\end{appendix}

\end{document}